\documentclass[aps,preprint]{revtex4}

\usepackage{amsfonts}
\usepackage{amsmath}
\usepackage{amssymb}
\usepackage[dvips]{graphicx}
\usepackage{bm}

\newtheorem{definition}{Definition}[section]
\newtheorem{theorem}[definition]{Theorem}
\newenvironment{proof}[1][Proof]{\noindent\textbf{#1.} }{\ \rule{0.5em}{0.5em}}

\begin{document}

\title{Expanding Lie (super)algebras through abelian semigroups}

\date{May 26, 2006}

\author{Fernando Izaurieta}
\email{fizaurie@gmail.com}

\author{Eduardo Rodr\'{\i}guez}
\email{edurodriguez@udec.cl}

\author{Patricio Salgado}
\email{pasalgad@udec.cl}

\affiliation{Departamento de F\'{\i}sica, Universidad de Concepci\'{o}n, Casilla 160-C, Concepci\'{o}n, Chile}

\begin{abstract}
We propose an outgrowth of the expansion method introduced by de~Azc\'{a}rraga \textit{et al.} [Nucl. Phys. B~\textbf{662} (2003) 185]. The basic idea consists in considering the direct product between an abelian semigroup $S$ and a Lie algebra $\mathfrak{g}$. General conditions under which relevant subalgebras can systematically be extracted from $S \times \mathfrak{g}$ are given. We show how, for a particular choice of semigroup $S$, the known cases of expanded algebras can be reobtained, while new ones arise from different choices. Concrete examples, including the M~algebra and a D'Auria--Fr\'{e}-like Superalgebra, are considered. Finally, we find explicit, non-trace invariant tensors for these $S$-expanded algebras, which are essential ingredients in, e.g., the formulation of Supergravity theories in arbitrary space-time dimensions.
\end{abstract}

\preprint{GACG/04/2006}

\maketitle

\tableofcontents

\section{Introduction}

The r\^{o}le played by Lie algebras and their interrelations in physics can hardly be over\-emphasized. To mention only one example, the Poincar\'{e} algebra may be obtained from the Galilei algebra via a \emph{deformation} process. This deformation is one of the ways in which different Lie algebras can be related.

The purpose of this paper is to shed new light on the method of \emph{expansion} of Lie algebras (for a thorough treatment, see the seminal work~\cite{Azcarraga Et Al} and references therein; early work on the subject is found in~\cite{Hatsuda Sakaguchi}). An Expansion is, in general, an algebra dimension-changing process. For instance, the M~algebra~\cite{Towsend1,Towsend2,CECS M}, with 583 Bosonic generators, can be regarded as an expansion of the orthosymplectic algebra $\mathfrak{osp} \left( 32|1 \right)$, which possesses only 528. This vantage viewpoint may help better understand fundamental problems related to the geometrical formulation of 11-dimensional Supergravity. Some physical applications of the expansion procedure have been developed in~\cite{Sak06,Azc05,Ban04,Ban04b,Hat04,Ban03,Hat03,Mee03,Iza06c}.

The approach to be presented here is entirely based on operations performed directly on the algebra generators, and thus differs from the outset with the one found in~\cite{Azcarraga Et Al}, where the dual Maurer--Cartan formulation is used. As a consequence, the expansion of free differential algebras lies beyond the scope of our analysis.

Finite abelian semigroups play a prominent r\^{o}le in our construction. All expansion cases found in~\cite{Azcarraga Et Al} may be regarded as coming from one particular choice of semigroup in the present approach, which is, in this sense, more general. Different semigroup choices yield, in general, expanded algebras that cannot be obtained by the methods of~\cite{Azcarraga Et Al}.

The plan of the paper goes as follows. After some preliminaries in section~\ref{sec:pre}, section~\ref{SecSExpPro} introduces the general procedure of abelian semigroup expansion, $S$-expansion for short, and shows how the cases found in~\cite{Azcarraga Et Al} can be recovered by an appropriate choice of semigroup $S$. In section~\ref{SecSExpSubAlg}, general conditions are given under which relevant subalgebras can be extracted from an $S$-expanded algebra. The case when $\mathfrak{g}$ satisfies the Weimar-Woods conditions~\cite{W-W,W-W2} and the case when $\mathfrak{g}$ is a superalgebra are studied. Section~\ref{SecSExpExamples} gives three explicit examples of $S$-expansions of $\mathfrak{osp}\left(  32|1\right)  $: (i)~the M~algebra~\cite{Towsend1,Towsend2,CECS M}, (ii)~a D'Auria--Fr\'{e}-like Superalgebra~\cite{DAuria-Fre} and (iii)~a new Superalgebra, different from but resembling aspects of the M~algebra, $\mathfrak{osp}\left(  32|1\right)  \times\mathfrak{osp}\left(  32|1\right)  $ and D'Auria--Fr\'{e} superalgebras. In section~\ref{SecOther}, the remaining cases of expanded algebras shown in~\cite{Azcarraga Et Al} are seen to also fit within the current scheme. The five-brane Superalgebra~\cite{vanHo82,deAz89} is given as an example. Section~\ref{SecInvTen} deals with the crucial problem of finding invariant tensors for the $S$-expanded algebras. General theorems are proven, allowing for nontrivial invariant tensors to be systematically constructed. We close in section~\ref{SecConclusions} with conclusions and an outlook for future work.

\section{\label{sec:pre}Preliminaries}

Before analyzing the $S$-Expansion procedure itself, it will prove convenient to introduce some basic notation and definitions.

\subsection{Semigroups}

\begin{definition}
Let $S = \left\{ \lambda_{\alpha} \right\}$ be a finite semigroup~\footnote{There does not seem to be a unique, universally-accepted definition of semigroup. Here it is taken to be a set provided with a closed associative product. It does not need to have an identity.}, and let us write the product of $\lambda_{\alpha_{1}}, \ldots, \lambda_{\alpha_{n}} \in S$ as
\begin{equation}
\lambda_{\alpha_{1}} \cdots \lambda_{\alpha_{n}} = \lambda_{\gamma \left( \alpha_{1}, \ldots, \alpha_{n} \right)}. \label{Lambda1x...xLambda n = Lambda Gamma}
\end{equation}
The \emph{$n$-selector} $K_{\alpha_{1} \cdots \alpha_{n}}^{\phantom{\alpha_{1} \cdots \alpha_{n}} \rho}$ is defined as
\begin{equation}
K_{\alpha_{1} \cdots \alpha_{n}}^{\phantom{\alpha_{1} \cdots \alpha_{n}} \rho} = \left\{
\begin{array}[c]{l}
1\text{, when }\rho=\gamma\left(  \alpha_{1},\ldots,\alpha_{n}\right)  \\
0\text{, otherwise.}%
\end{array}
\right.  \label{n-Selector}%
\end{equation}

Since $S$ is associative, the $n$-selector fulfills the identity
\begin{equation}
K_{\alpha_{1}\cdots\alpha_{n}}^{\phantom{\alpha_{1}\cdots\alpha_{n}}\rho} = K_{\alpha_{1}\cdots\alpha_{n-1}}^{\phantom{\alpha_{1}\cdots\alpha_{n-1}}\sigma} K_{\sigma\alpha_{n}}^{\phantom{\sigma\alpha_{n}}\rho} = K_{\alpha_{1}\sigma}^{\phantom{\alpha_{1}\sigma}\rho} K_{\alpha_{2}\cdots\alpha_{n}}^{\phantom{\alpha_{2}\cdots\alpha_{n}}\sigma}. \label{idsel}
\end{equation}
\end{definition}

Using this identity it is always possible to express the $n$-selector in terms of 2-selectors, which encode the information from the multiplication table of $S$.

An interesting way to state the same is that $2$-selectors provide a matrix
representation for $S$. As a matter of fact, when we write%
\begin{equation}
\left[ \lambda_{\alpha} \right]_{\mu}^{\phantom{\mu} \nu} =
K_{\mu \alpha}^{\phantom{\mu \alpha} \nu}
\label{MatrixSemigroupRep}%
\end{equation}
then we have%
\begin{equation}
\left[ \lambda_{\alpha} \right]_{\mu}^{\phantom{\mu} \sigma} \left[ \lambda_{\beta} \right]_{\sigma}^{\phantom{\sigma} \nu} = K_{\alpha \beta}^{\phantom{\alpha \beta} \sigma} \left[ \lambda_{\sigma} \right]_{\mu}^{\phantom{\mu} \nu} = \left[ \lambda_{\gamma \left( \alpha, \beta \right)} \right]_{\mu}^{\phantom{\mu} \nu}.
\end{equation}

We will restrict ourselves from now on to \emph{abelian} semigroups, which
implies that the $n$-selectors will be completely symmetrical in their lower indices.

The following definition introduces a product between semigroup subsets which will be extensively used throughout the paper.

\begin{definition}
\label{DefSubsetsProduct}
Let $S_{p}$ and $S_{q}$ be two subsets of $S$. The product $S_{p} \cdot S_{q}$  is defined as
\begin{equation}
S_{p}\cdot S_{q}=\left\{  \lambda_{\gamma}\text{ such that }\lambda_{\gamma}=\lambda_{\alpha_{p}}\lambda_{\alpha_{q}}\text{, with }\lambda_{\alpha_{p}} \in S_{p}\text{ and }\lambda_{\alpha_{q}}\in S_{q}\right\}  \subset S. \label{Subsets Product}
\end{equation}
In other words, $S_{p}\cdot S_{q}\subset S$ is the set which results from the product of every element of $S_{p}$ with every element of $S_{q}$. Since $S$ is abelian, $S_{p}\cdot S_{q}=S_{q}\cdot S_{p}$.
\end{definition}

Let us emphasize that, in general, $S_{p}$, $S_{q}$ and $S_{p}\cdot S_{q}$
need \emph{not} be semigroups by themselves.

The abelian semigroup $S$ could also be provided with a unique zero element, $0_{S}$. This element is defined as the one for which
\begin{equation}
0_{S}\lambda_{\alpha}=\lambda_{\alpha}0_{S}=0_{S},
\end{equation}
for each $\lambda_{\alpha}\in S$.

\subsection{Reduced Lie Algebras}

The following definition introduces the concept of \emph{reduction} of Lie algebras.

\begin{definition}
\label{def:foralg}
Consider a Lie (super)algebra $\mathfrak{g}$ of the form $\mathfrak{g}%
=V_{0}\oplus V_{1}$, with $\left\{  \bm{T}_{a_{0}}\right\}  $ being a basis for $V_{0}$ and $\left\{  \bm{T}_{a_{1}}\right\}  $ a basis for $V_{1}.$ When $\left[  V_{0},V_{1}\right]  \subset V_{1}$, i.e., when the commutation relations have the general form%
\begin{eqnarray}
\left[ \bm{T}_{a_{0}}, \bm{T}_{b_{0}} \right] & = & C_{a_{0} b_{0}}^{\phantom{a_{0} b_{0}} c_{0}} \bm{T}_{c_{0}} + C_{a_{0} b_{0}}^{\phantom{a_{0} b_{0}} c_{1}} \bm{T}_{c_{1}}, \label{ReducedAlgebra 00} \\
\left[ \bm{T}_{a_{0}}, \bm{T}_{b_{1}} \right] & = & C_{a_{0} b_{1}}^{\phantom{a_{0} b_{1}} c_{1}} \bm{T}_{c_{1}}, \label{ReducedAlgebra 01} \\
\left[ \bm{T}_{a_{1}}, \bm{T}_{b_{1}} \right] & = & C_{a_{1} b_{1}}^{\phantom{a_{1} b_{1}} c_{0}} \bm{T}_{c_{0}} + C_{a_{1} b_{1}}^{\phantom{a_{1} b_{1}} c_{1}} \bm{T}_{c_{1}}, \label{ReducedAlgebra 11}
\end{eqnarray}
then it is straightforward to show that the structure constants $C_{a_{0} b_{0}}^{\phantom{a_{0} b_{0}} c_{0}}$ satisfy the Jacobi identity by themselves, and therefore $\left[ \bm{T}_{a_{0}}, \bm{T}_{b_{0}} \right] = C_{a_{0} b_{0}}^{\phantom{a_{0} b_{0}} c_{0}} \bm{T}_{c_{0}}$ corresponds by itself to a Lie (super)algebra. This algebra, with structure constants $C_{a_{0} b_{0}}^{\phantom{a_{0} b_{0}} c_{0}}$, is called a \emph{reduced algebra} of $\mathfrak{g}$ and symbolized as $\left\vert V_{0}\right\vert .$
\end{definition}

The reduced algebra could be regarded in some way as the ``inverse'' of an
algebra extension, but $V_{1}$ does not need to be an ideal. Note also that a reduced algebra \emph{does not} in general correspond to a subalgebra.

\section{\label{SecSExpPro}The $S$-Expansion Procedure}

\subsection{$S$-Expansion for an Arbitrary Semigroup $S$}

The following theorem embodies one of the main results of the paper, the concept of $S$-expanded algebras.

\begin{theorem}
\label{S-ExpanTh}Let $S=\left\{  \lambda_{\alpha}\right\}  $ be an abelian
semigroup with $2$-selector $K_{\alpha \beta}^{\phantom{\alpha \beta} \gamma}$ and $\mathfrak{g}$ a Lie (super)algebra with basis $\left\{
\bm{T}_{A}\right\}  $ and structure constants $C_{AB}^{\phantom{AB} C}$. Denote a basis element of the direct product $S \times \mathfrak{g}$ by $\bm{T}_{\left(  A,\alpha\right)  }=\lambda_{\alpha}\bm{T}_{A}$ and consider the induced commutator $\left[ \bm{T}_{\left(  A,\alpha\right)  },\bm{T}_{\left(
B,\beta\right)  }\right]  \equiv\lambda_{\alpha}\lambda_{\beta}\left[
\bm{T}_{A},\bm{T}_{B}\right]  $. Then, $S\times\mathfrak{g}$ is also a Lie (super)algebra with structure constants
\begin{equation}
C_{\left( A, \alpha \right) \left( B, \beta \right)}^{\phantom{\left( A, \alpha \right) \left( B, \beta \right)} \left( C, \gamma \right)} = K_{\alpha \beta}^{\phantom{\alpha \beta} \gamma} C_{AB}^{\phantom{AB} C}. \label{C=KC}%
\end{equation}
\end{theorem}

\begin{proof}
Starting from the form of the induced commutator and using the multiplication
law~(\ref{Lambda1x...xLambda n = Lambda Gamma}) one finds%
\begin{eqnarray*}
\left[  \bm{T}_{\left(  A,\alpha\right)  },\bm{T}_{\left(
B,\beta\right)  }\right]   &  \equiv & \lambda_{\alpha}\lambda_{\beta}\left[
\bm{T}_{A},\bm{T}_{B}\right]  \\
&  = & C_{AB}^{\phantom{AB} C} \lambda_{\gamma\left(  \alpha,\beta\right)
}\bm{T}_{C}\\
&  = & C_{AB}^{\phantom{AB} C} \bm{T}_{\left(  C,\gamma\left(
\alpha,\beta\right)  \right)  }.
\end{eqnarray*}
The definition of the 2-selector $K_{\alpha \beta}^{\phantom{\alpha \beta} \rho}$ [cf.~eq.~(\ref{n-Selector})],
\[
K_{\alpha \beta}^{\phantom{\alpha \beta} \rho}=\left\{
\begin{array}[c]{l}
1\text{, when }\rho=\gamma\left(  \alpha,\beta\right)  \\
0\text{, otherwise,}%
\end{array}
\right.
\]
now allows us to write%
\begin{equation}
\left[  \bm{T}_{\left(  A,\alpha\right)  },\bm{T}_{\left(
B,\beta\right)  }\right] = K_{\alpha \beta}^{\phantom{\alpha \beta} \rho} C_{AB}^{\phantom{AB} C}\bm{T}_{\left(  C,\rho\right)  }.
\end{equation}
Therefore, the algebra spanned by $\left\{  \bm{T}_{\left(
A,\alpha\right)  }\right\}  $ closes and the structure constants read%
\begin{equation}
C_{\left( A, \alpha \right) \left( B, \beta \right)}^{\phantom{\left( A, \alpha \right) \left( B, \beta \right)} \left( C, \gamma \right)} = K_{\alpha \beta}^{\phantom{\alpha \beta} \gamma} C_{AB}^{\phantom{AB} C}.
\end{equation}

Since $S$ is abelian, the structure constants $C_{\left( A, \alpha \right) \left( B, \beta \right)}^{\phantom{\left( A, \alpha \right) \left( B, \beta \right)} \left( C, \gamma \right)}$ have the same symmetries as $C_{AB}^{\phantom{AB} C}$, namely
\begin{equation}
C_{\left( A, \alpha \right) \left( B, \beta \right)}^{\phantom{\left( A, \alpha \right) \left( B, \beta \right)} \left( C, \gamma \right)} = - \left( -1 \right)^{\mathfrak{q} \left( A \right) \mathfrak{q} \left( B \right)} C_{ \left( B, \beta \right) \left( A, \alpha \right)}^{\phantom{\left( A, \alpha \right) \left( B, \beta \right)} \left( C, \gamma \right)},
\end{equation}
and for this reason, $\mathfrak{q} \left(  A,\alpha \right)
=\mathfrak{q}\left(  A\right)  $, where $\mathfrak{q}\left(  A\right)  $
denotes the degree of $\bm{T}_{A}$ (1 for Fermi and 0 for Bose).

In order to show that the structure constants $C_{\left( A, \alpha \right) \left( B, \beta \right)}^{\phantom{\left( A, \alpha \right) \left( B, \beta \right)} \left( C, \gamma \right)}$ satisfy the Jacobi identity, it suffices to use the properties of the selectors [cf.~eq.~(\ref{idsel})] and the fact that the structure constants $C_{AB}^{\phantom{AB} C}$ satisfy the Jacobi identity themselves. This concludes the proof.
\end{proof}

The following definition is a natural outcome of Theorem~\ref{S-ExpanTh}.

\begin{definition}
Let $S$ be an abelian semigroup and $\mathfrak{g}$ a Lie algebra. The Lie algebra $\mathfrak{G}$ defined by $\mathfrak{G}=S\times\mathfrak{g}$ is
called \emph{$S$-Expanded algebra} of $\mathfrak{g}$.
\end{definition}

When the semigroup has a zero element $0_{S} \in S$, it plays a somewhat peculiar r\^{o}le in the $S$-expanded algebra. Let us span $S$ in nonzero elements $\lambda
_{i},$ $i=0,\ldots,N$ and a zero element $\lambda_{N+1}=0_{S}$. Then, the
$2$-selector satisfies%
\begin{eqnarray}
K_{i,N+1}^{\phantom{i,N+1} j} & = & K_{N+1,i}^{\phantom{N+1,i} j}=0,\\
K_{i,N+1}^{\phantom{i,N+1} N+1} & = & K_{N+1,i}^{\phantom{N+1,i} N+1}=1,\\
K_{N+1,N+1}^{\phantom{N+1,N+1} j} & = & 0,\\
K_{N+1,N+1}^{\phantom{N+1,N+1} N+1} & = & 1.
\end{eqnarray}

Therefore, $\mathfrak{G}=S\times\mathfrak{g}$ can be split as
\begin{eqnarray}
\left[  \bm{T}_{\left(  A,i\right)  },\bm{T}_{\left(
B,j\right)  }\right]   &  = & K_{ij}^{\phantom{ij} k} C_{AB}^{\phantom{AB} C} \bm{T}_{\left(  C,k\right)  }+K_{ij}^{\phantom{ij} N+1}C_{AB}^{\phantom{AB} C} \bm{T}_{\left(  C,N+1\right)  }, \label{TTij} \\
\left[  \bm{T}_{\left(  A,N+1\right)  },\bm{T}_{\left(
B,j\right)  }\right]   &  = & C_{AB}^{\phantom{AB} C} \bm{T}_{\left(
C,N+1\right)  }, \label{TTNj} \\
\left[  \bm{T}_{\left(  A,N+1\right)  },\bm{T}_{\left(
B,N+1\right)  }\right]   &  = & C_{AB}^{\phantom{AB} C} \bm{T}_{\left(
C,N+1\right)  }. \label{TTNN}
\end{eqnarray}

Comparing~(\ref{TTij})--(\ref{TTNN}) with~(\ref{ReducedAlgebra 00})--(\ref{ReducedAlgebra 11}), one sees that the commutation relations
\begin{equation}
\left[  \bm{T}_{\left(  A,i\right)  },\bm{T}_{\left(
B,j\right)  }\right]  =K_{ij}^{\phantom{ij} k}C_{AB}^{\phantom{AB} C}\bm{T}_{\left(  C,k\right)  } \label{0-Reduced Algebra}
\end{equation}
are those of a \emph{reduced} Lie algebra of $\mathfrak{G}$ (see Def.~\ref{def:foralg}). The reduction procedure in this particular case is equivalent to imposing the condition
\begin{equation}
\bm{T}_{\left( A, N+1 \right) } = 0_{S} \bm{T}_{A} = \bm{0}. \label{0sxT=0}
\end{equation}

Notice that in this case the reduction abelianizes large sectors of the algebra; for each $i$ and $j$ satisfying $K_{ij}^{\phantom{ij} N+1}=1$ (i.e., $\lambda_{i} \lambda_{j} = \lambda_{N+1}$) we have $\left[ \bm{T}_{\left( A, i \right)}, \bm{T}_{\left( B, j \right)} \right] = \bm{0}$.

The above considerations motivate the following definition:

\begin{definition}
Let $S$ be an abelian semigroup with a zero element $0_{S} \in S$, and let
$\mathfrak{G} = S \times \mathfrak{g}$ be an $S$-expanded algebra. The algebra obtained by imposing the condition $0_{S} \bm{T}_{A} = \bm{0}$ on $\mathfrak{G}$ (or a subalgebra of it) is called \emph{$0_{S}$-reduced algebra} of $\mathfrak{G}$ (or of the subalgebra).
\end{definition}

The algebra~(\ref{0-Reduced Algebra}) appears naturally when the semigroup's zero matches the (algebra) field's zero. As we will see in the next section, this is the way Maurer--Cartan forms power-series expanded algebras fit within the present scheme. It is also possible to extract other reduced algebras from $\mathfrak{G}$; as will be analyzed in Sec.~\ref{SecOther}, the $0_{S}$-reduced algebra turns out to be a particular case of Theorem~\ref{ReducedAlgTh}.

\subsection{\label{SecAlgExpSExp}Maurer--Cartan Forms Power Series Algebra Expansion as an $S$-Expansion}

The Maurer--Cartan forms power series algebra expansion method is a powerful
procedure which can lead, in stark contrast with contraction, deformation and
extension of algebras, to algebras of a dimension higher than the original
one. In a nutshell, the idea consists of looking at the algebra $\mathfrak{g}$
as described by the associated Maurer--Cartan forms on the group manifold and,
after rescaling some of the group parameters by a factor $\lambda$, in
expanding the Maurer--Cartan forms as a power series in $\lambda$. Finally
this series is truncated in a way that assures the closure of the algebra. The
subject is thoroughly treated by de~Azc\'{a}rraga and Izquierdo in
Ref.~\cite{LibroAzcarraga} and de~Azc\'{a}rraga, Izquierdo, Pic\'{o}n and
Varela in Ref.~\cite{Azcarraga Et Al}.

Theorem 1 of Ref.~\cite{Azcarraga Et Al} shows that, in the more general case, the expanded Lie algebra has the structure constants
\begin{equation}
C_{\left( A, i \right) \left( B, j \right)}^{\phantom{\left( A, i \right) \left( B, j \right)} \left( C, k \right)} = \left\{
\begin{array}[c]{cl}
0, & \text{when } i+j \neq k \\
C_{AB}^{\phantom{AB} C}, & \text{when } i+j=k
\end{array}
\right.  , \label{scazc}
\end{equation}
where the parameters $i,j,k=0,\ldots,N$ correspond to the order of the
expansion, and $N$ is the truncation order.

These structure constants can also be obtained within the $S$-expansion procedure. In order to show this, one must consider the $0_{S}$-reduction of an $S$-expanded algebra where $S$ corresponds to the semigroup defined below.

\begin{definition}
Let us define $S_{\mathrm{E}}^{\left( N \right)}$ as the semigroup of elements
\begin{equation}
S_{\mathrm{E}}^{\left( N \right)} = \left\{ \lambda_{\alpha}, \alpha = 0, \ldots, N+1 \right\}, \label{SEN}
\end{equation}
provided with the multiplication rule
\begin{equation}
\lambda_{\alpha} \lambda_{\beta} = \lambda_{H_{N+1}\left( \alpha + \beta \right)}, \label{Normal Expansion Product}
\end{equation}
where $H_{N+1}$ is defined as the function
\begin{equation}
H_{n} \left( x \right) = \left\{
\begin{array}[c]{l}
x \text{, when } x < n \\
n \text{, when } x \geq n
\end{array}
\right.  .
\end{equation}
The $2$-selectors for $S_{\mathrm{E}}^{\left(  N\right)  }$ read
\begin{equation}
K_{\alpha \beta}^{\phantom{\alpha \beta} \gamma} = \delta_{H_{N+1}\left(  \alpha+\beta\right)}^{\gamma},
\end{equation}
where $\delta_{\sigma}^{\rho}$ is the Kronecker delta. From eq.~(\ref{Normal Expansion Product}), we have that $\lambda_{N+1}$ is the zero element in $S_{\mathrm{E}}^{\left(  N\right)  }$, i.e., $\lambda_{N+1}=0_{S}$.
\end{definition}

Using eq.~(\ref{C=KC}), the structure constants for the $S_{\mathrm{E}}^{\left( N \right)}$-expanded algebra can be written as
\begin{equation}
C_{\left( A, \alpha \right) \left( B, \beta \right)}^{\phantom{\left( A, \alpha \right) \left( B, \beta \right)} \left( C, \gamma \right)} = \delta_{H_{N+1} \left( \alpha + \beta \right)}^{\gamma} C_{AB}^{\phantom{AB} C}, \label{C=DeltaN+1C}%
\end{equation}
with $\alpha,\beta,\gamma=0,\ldots,N+1$. When the extra condition $\lambda_{N+1}\bm{T}%
_{A}=\bm{0}$ is imposed, eq.~(\ref{C=DeltaN+1C}) reduces to%
\begin{equation}
C_{\left( A, i \right) \left( B, j \right)}^{\phantom{\left( A, i \right) \left( B, j \right)} \left( C, k \right)} = \delta_{i+j}^{k} C_{AB}^{\phantom{AB} C}, \label{No le quiero poner Numero}
\end{equation}
which exactly matches the structure constants~(\ref{scazc}).

The above arguments show that the Maurer--Cartan forms power series expansion of an algebra $\mathfrak{g}$, with truncation order $N$, coincides with the $0_{S}$-reduction of the $S_{\mathrm{E}}^{(N)}$-expanded algebra $\mathfrak{G}^{\left( \mathrm{E} \right)} = S_{\mathrm{E}}^{(N)} \times \mathfrak{g}$.

This is of course no coincidence. The set of powers of the rescaling parameter
$\lambda$, together with the truncation at order $N$, satisfy precisely the
multiplication law of $S_{\mathrm{E}}^{\left(  N\right)  } $. As a
matter of fact, we have%
\begin{equation}
\lambda^{\alpha}\lambda^{\beta}=\lambda^{\alpha+\beta},
\end{equation}
and the truncation can be imposed as%
\begin{equation}
\lambda^{\alpha}=0\text{ when }\alpha>N.
\end{equation}

It is for this reason that one must demand $0_{S}\bm{T}_{A}%
=\bm{0}$ in order to obtain the MC~expansion as an $S_{\mathrm{E}}$-expansion: in this case the zero of the semigroup is the zero of the field
as well.

The $S$-expansion procedure is valid no matter what the structure of the
original Lie algebra $\mathfrak{g}$ is, and in this sense it is very general.
However, when something about the structure of $\mathfrak{g}$ is known, a lot
more can be done. As an example, in the context of MC~expansion, the rescaling
and truncation can be performed in several ways depending on the structure of
$\mathfrak{g}$, leading to several kinds of expanded algebras. Important
examples of this are the generalized \.{I}n\"{o}n\"{u}--Wigner contraction, or
the M~algebra as an expansion of $\mathfrak{osp}\left(  32|1\right)  $ (see
Refs.~\cite{Azcarraga Et Al,AzcarragaSuperspace}). This is also the case in
the context of $S$-expansions. As we will show in the next section, when some information about the structure of $\mathfrak{g}$ is available, it is possible to find subalgebras of $\mathfrak{G}=S\times\mathfrak{g}$ and other kinds of reduced algebras. In this way, all the algebras obtained by the MC~expansion procedure can be reobtained. New kinds of $S$-expanded algebras can also be obtained by considering semigroups different from $S_{\mathrm{E}}$.

\section{\label{SecSExpSubAlg}$S$-Expansion Subalgebras}

An $S$-expanded algebra has a fairly simple structure. In a way, it reproduces the original algebra $\mathfrak{g}$ in a series of ``levels'' corresponding to the semigroup elements. Interestingly enough, there are at least two ways of extracting smaller algebras from $S \times \mathfrak{g}$. The first one, described in this section, gives rise to a ``resonant subalgebra,'' while the second, described in section~\ref{SecOther}, produces reduced algebras (in the sense of Def.~\ref{def:foralg}).

\subsection{Resonant Subalgebras for an Arbitrary Semigroup $S$}

The general problem of finding subalgebras from an $S$-expanded algebra is a nontrivial one, which is met and solved (in a particular setting) in this section (see theorem~\ref{ResSubAlgTh} below). In order to provide a solution, one must have some information on the subspace structure of $\mathfrak{g}$. This information is encoded in the following way.

Let $\mathfrak{g} = \bigoplus_{p\in I} V_{p}$ be a decomposition of $\mathfrak{g}$ in subspaces $V_{p}$, where $I$ is
a set of indices. For each $p,q\in I$ it is always possible to define
$i_{\left(  p,q\right)  }\subset I$ such that%
\begin{equation}
\left[ V_{p}, V_{q} \right] \subset \bigoplus_{r \in i_{\left( p, q \right) }} V_{r}. \label{VpVq=Vr}
\end{equation}
In this way, the subsets $i_{\left(  p,q\right)  }$ store the information on the subspace structure of $\mathfrak{g}$.

As for the abelian semigroup $S$, this can always be decomposed as $S = \bigcup_{p \in I} S_{p}$, where $S_{p} \subset S$. In principle, this decomposition is completely arbitrary; however, using the product from Def.~\ref{DefSubsetsProduct}, it is sometimes possible to pick up a very particular choice of subset decomposition. This choice is the subject of the following

\begin{definition}
Let $\mathfrak{g} = \bigoplus_{p \in I} V_{p}$ be a decomposition of $\mathfrak{g}$ in subspaces, with a structure described by the subsets $i_{\left( p, q \right)}$, as in eq.~(\ref{VpVq=Vr}). Let $S = \bigcup_{p \in I} S_{p}$ be a subset decomposition of the abelian semigroup $S$ such that
\begin{equation}
S_{p} \cdot S_{q} \subset \bigcap_{r \in i_{\left( p, q \right)}} S_{r}, \label{SpSq=Sr}
\end{equation}
where the subset product $\cdot$ is the one from Def.~\ref{DefSubsetsProduct}. When such subset decomposition $S = \bigcup_{p \in I} S_{p}$ exists, then we say that this decomposition is in \emph{resonance} with the subspace decomposition of $\mathfrak{g}$, $\mathfrak{g} = \bigoplus_{p \in I} V_{p}.$
\end{definition}

The resonant subset decomposition is crucial in order to systematically extract subalgebras from the $S$-expanded algebra $\mathfrak{G} = S \times \mathfrak{g},$ as is proven in the following

\begin{theorem}
\label{ResSubAlgTh}Let $\mathfrak{g} = \bigoplus_{p \in I} V_{p}$ be a subspace decomposition of $\mathfrak{g}$, with a structure described by eq.~(\ref{VpVq=Vr}), and let $S = \bigcup_{p \in I} S_{p}$ be a resonant subset decomposition of the abelian semigroup $S$, with the structure given in eq.~(\ref{SpSq=Sr}). Define the subspaces of $\mathfrak{G} = S \times \mathfrak{g},$
\begin{equation}
W_{p}=S_{p}\times V_{p}, \qquad p \in I.
\end{equation}
Then,
\begin{equation}
\mathfrak{G}_{\mathrm{R}} = \bigoplus_{p \in I} W_{p}
\end{equation}
is a subalgebra of $\mathfrak{G} = S \times \mathfrak{g}$.
\end{theorem}

\begin{proof}
Using eqs.~(\ref{VpVq=Vr})--(\ref{SpSq=Sr}), we have
\begin{eqnarray}
\left[ W_{p}, W_{q} \right] & \subset & \left( S_{p} \cdot S_{q} \right) \times \left[ V_{p}, V_{q} \right] \nonumber \\
& \subset & \bigcap_{s \in i_{\left( p, q \right)}} S_{s} \times \bigoplus_{r \in i_{\left( p, q \right)}} V_{r} \nonumber \\
& \subset & \bigoplus_{r \in i_{\left( p, q \right)}} \left[ \bigcap_{s \in i_{\left( p, q \right)}} S_{s} \right] \times V_{r}
\end{eqnarray}

Now, it is clear that for each $r \in i_{\left( p, q \right)}$, one can write
\begin{equation}
\bigcap_{s \in i_{\left( p, q \right)}} S_{s} \subset S_{r} .
\end{equation}

Then,
\begin{equation}
\left[ W_{p}, W_{q} \right] \subset \bigoplus_{r \in i_{\left( p, q \right)}} S_{r} \times V_{r}
\end{equation}
and we arrive at%
\begin{equation}
\left[ W_{p}, W_{q} \right] \subset \bigoplus_{r \in i_{\left( p, q \right)}} W_{r}. \label{[Wp,Wq]=+Wr}
\end{equation}

Therefore, the algebra closes and $\mathfrak{G}_{\mathrm{R}} = \bigoplus_{p \in I} W_{p}$ is a subalgebra of $\mathfrak{G}$.
\end{proof}

\begin{definition}
The algebra $\mathfrak{G}_{\mathrm{R}} = \bigoplus_{p \in I} W_{p}$ obtained in Theorem~\ref{ResSubAlgTh} is called a \emph{Resonant Subalgebra} of the $S$-expanded algebra $\mathfrak{G} = S \times \mathfrak{g}$.
\end{definition}

The choice of the name resonance is due to the formal similarity between eqs.~(\ref{VpVq=Vr}) and~(\ref{SpSq=Sr}); eq.~(\ref{SpSq=Sr}) will be also referred to as ``\emph{resonance condition}.''

Theorem~\ref{ResSubAlgTh} translates the difficult problem of finding subalgebras from an $S$-expanded algebra $\mathfrak{G} = S \times \mathfrak{g}$ into that of finding a resonant partition for the semigroup $S$. As the examples from section~\ref{SecSExpExamples} help make clear, solving the resonance condition~(\ref{SpSq=Sr}) turns out to be an easily tractable problem. Theorem~\ref{ResSubAlgTh} can thus be regarded as a useful tool for extracting subalgebras from an $S$-expanded algebra.

Using eq.~(\ref{C=KC}) and the resonant subset partition of $S$ it is possible
to find an explicit expression for the structure constants of the resonant
subalgebra $\mathfrak{G}_{\mathrm{R}}$. Denoting the basis of $V_{p}$ by $\left\{ \bm{T}_{a_{p}} \right\}$, one can write
\begin{equation}
C_{\left( a_{p}, \alpha_{p} \right) \left( b_{q}, \beta_{q} \right)}^{\phantom{\left( a_{p}, \alpha_{p} \right) \left( b_{q}, \beta_{q} \right)} \left( c_{r}, \gamma_{r} \right)} = K_{\alpha_{p} \beta_{q}}^{\phantom{\alpha_{p} \beta_{q}} \gamma_{r}} C_{a_{p} b_{q}}^{\phantom{a_{p} b_{q}} c_{r}} \text{ with } \alpha_{p}, \beta_{q}, \gamma_{r} \text{ such that } \lambda_{\alpha_{p}} \in S_{p},\lambda_{\beta_{q}} \in S_{q}, \lambda_{\gamma_{r}} \in S_{r} . \label{ResonantStructureConstants}
\end{equation}

An interesting fact is that the $S$-expanded algebra ``subspace structure''
encoded in $i_{\left(  p,q\right)  }$ is the same as in the original algebra
$\mathfrak{g}$, as can be observed from eq.~(\ref{[Wp,Wq]=+Wr}).

Resonant Subalgebras play a central r\^{o}le in the current scheme. It is
interesting to notice that most of the cases considered in~\cite{Azcarraga Et
Al} can be reobtained using the above theorem for $S=S_{\mathrm{E}}$
[recall eqs.~(\ref{SEN})--(\ref{Normal Expansion Product})] and $0_{S}$-reduction, as we will see in the next section. All remaining cases can be obtained as a more general reduction of a resonant subalgebra (see sec.~\ref{SecOther}).

\subsection{$S=S_{\mathrm{E}}$ Resonant Subalgebras and MC Expanded Algebras}

In this section, some results presented for algebra expansions in Ref.~\cite{Azcarraga Et Al} are recovered within the $S$-expansion approach. To get these algebras one must proceed in a three-step fashion:
\begin{enumerate}
\item Perform an $S$-expansion using the semigroup $S=S_{\mathrm{E}}$,
\item Find a resonant partition for $S_{\mathrm{E}}$ and construct the resonant subalgebra $\mathfrak{G}_{\mathrm{R}}$,
\item Apply a $0_{S}$-reduction (or a more general one, see sec.~\ref{SecOther}) to the resonant subalgebra.
\end{enumerate}

Choosing a different semigroup $S$ or omitting the reduction procedure one finds algebras not contained within the Maurer--Cartan forms power series expansion of Ref.~\cite{Azcarraga Et Al}. Such an example is provided in sec.~\ref{SecZ4Exp}.

\subsubsection{Case when $\mathfrak{g} = V_{0} \oplus V_{1}$, with $V_{0}$ being a Subalgebra and $V_{1}$ a Symmetric Coset}

Let $\mathfrak{g}=V_{0}\oplus V_{1}$ be a subspace decomposition of
$\mathfrak{g}$, such that
\begin{eqnarray}
\left[  V_{0},V_{0}\right]   &  \subset  & V_{0},\\
\left[  V_{0},V_{1}\right]   &  \subset  & V_{1},\\
\left[  V_{1},V_{1}\right]   &  \subset  & V_{0}.
\end{eqnarray}

Let $S_{\mathrm{E}}^{\left(  N\right)  }=S_{0}\cup S_{1}$, with $N$
arbitrary, be a subset decomposition of $S_{\mathrm{E}}$,
with~\footnote{Here $\left[  x\right]  $ denotes the integer part of $x$.}%
\begin{eqnarray}
S_{0}  &  = & \left\{  \lambda_{2m},\text{ with }m=0,\ldots,\left[  \frac{N}%
{2}\right]  \right\}  \cup\left\{  \lambda_{N+1}\right\}  ,\\
S_{1}  &  = & \left\{  \lambda_{2m+1},\text{ with }m=0,\ldots,\left[  \frac
{N-1}{2}\right]  \right\}  \cup\left\{  \lambda_{N+1}\right\}  .
\end{eqnarray}

This subset decomposition of $S_{\mathrm{E}}^{\left(  N\right)  }$
satisfies the resonance condition~(\ref{SpSq=Sr}), which in this case
explicitly reads
\begin{eqnarray}
S_{0}\cdot S_{0} &  \subset & S_{0}\\
S_{0}\cdot S_{1} &  \subset & S_{1}\\
S_{1}\cdot S_{1} &  \subset & S_{0}.
\end{eqnarray}

Therefore, according to Theorem~\ref{ResSubAlgTh}, we have that%
\begin{equation}
\mathfrak{G}_{\mathrm{R}}=W_{0}\oplus W_{1},
\end{equation}
with
\begin{eqnarray}
W_{0}  &  = & S_{0}\times V_{0},\\
W_{1}  &  = & S_{1}\times V_{1},
\end{eqnarray}
is a resonant subalgebra of $\mathfrak{G}.$

Using eq.~(\ref{ResonantStructureConstants}), it is straightforward to write the structure constants for the resonant subalgebra,
\[
C_{\left( a_{p}, \alpha_{p} \right) \left( b_{q}, \beta_{q} \right)}^{\phantom{\left( a_{p}, \alpha_{p} \right) \left( b_{q}, \beta_{q} \right)} \left( c_{r}, \gamma_{r} \right)} = \delta_{H_{N+1} \left( \alpha_{p} + \beta_{q} \right)}^{\gamma_{r}} C_{a_{p} b_{q}}^{\phantom{a_{p} b_{q}} c_{r}} \text{ with } \left\{
\begin{array}
[c]{l}%
p,q=0,1\\
\alpha_{p},\beta_{p},\gamma_{p}=2m+p, \\
m=0, \ldots,\left[  \frac{N-p}%
{2}\right]  ,\frac{N+1-p}{2}.
\end{array}
\right.
\]

Imposing $\lambda_{N+1}\bm{T}_{A}=\bm{0}$, the $0_{S}$-reduced structure constants are obtained as
\begin{equation}
C_{\left( a_{p}, \alpha_{p} \right) \left( b_{q}, \beta_{q} \right)}^{\phantom{\left( a_{p}, \alpha_{p} \right) \left( b_{q}, \beta_{q} \right)} \left( c_{r}, \gamma_{r} \right)} = \delta_{H_{N+1} \left( \alpha_{p} + \beta_{q} \right)}^{\gamma_{r}} C_{a_{p} b_{q}}^{\phantom{a_{p} b_{q}} c_{r}} \text{ with } \left\{
\begin{array}
[c]{l}%
p,q=0,1\\
\alpha_{p},\beta_{p},\gamma_{p}=2m+p, \\
m=0,\ldots,\left[  \frac{N-p}{2}\right]  .
\end{array}
\right.   \label{xxxStrucConstSymCoset}
\end{equation}
In order to compare with the MC Expansion, let us observe that, with the notation of~\cite{Azcarraga Et Al}, the $0_{S}$-reduction of the $S_{\mathrm{E}}$-expanded algebra corresponds to $\mathcal{G} \left( N_{0} , N_{1} \right)$ for the symmetric coset case, with
\begin{eqnarray}
N_{0}  & = & 2\left[  \frac{N}{2}\right]  ,\\
N_{1}  & = & 2\left[  \frac{N-1}{2}\right]  +1.
\end{eqnarray}

The structure constants~(\ref{xxxStrucConstSymCoset}) correspond to the structure constants~(3.31) from Ref.~\cite{Azcarraga Et Al} (the notation is slightly different though).

\begin{figure}
\includegraphics[width=\columnwidth]{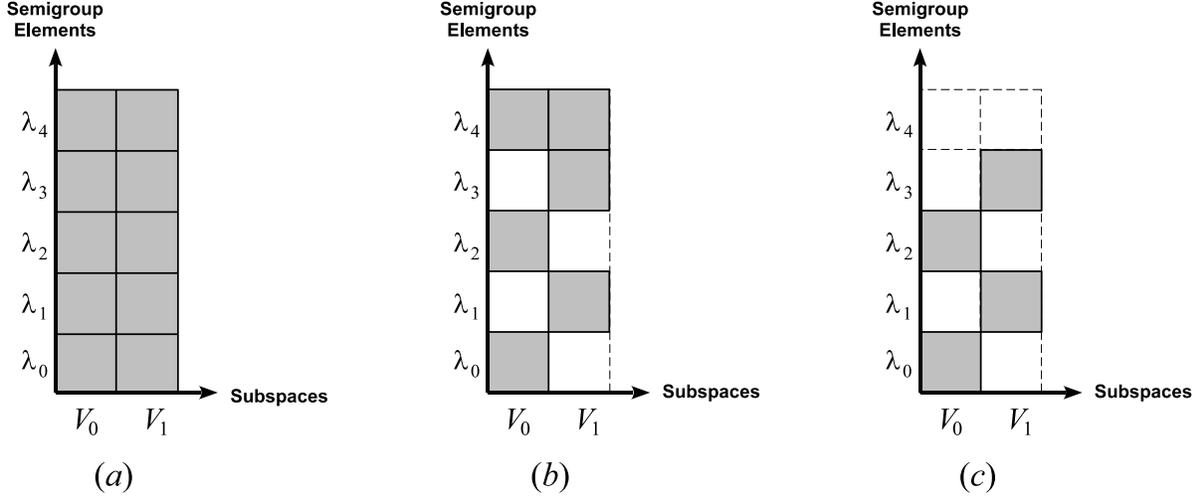}
\caption{\label{FigSymCosetRes}$S_{\mathrm{E}}^{(3)}$-expansion of an algebra $\mathfrak{g} = V_{0} \oplus V_{1}$, where $V_{0}$ is a subalgebra and $V_{1}$ a symmetric coset. (\textit{a}) The gray region corresponds to the full $S_{\mathrm{E}}^{(3)}$-expanded algebra, $\mathfrak{G} = S_{\mathrm{E}}^{(3)} \times \mathfrak{g}$. (\textit{b}) The shaded area here depicts a resonant subalgebra $\mathfrak{G}_{\mathrm{R}}$. (\textit{c}) The gray region now shows the $0_{S}$-reduction of the resonant subalgebra $\mathfrak{G}_{\mathrm{R}}$.}
\end{figure}

A more intuitive idea of the whole procedure of $S$-expansion, resonant subalgebra and $0_{S}$-reduction can be obtained by means of a diagram, such as the one depicted in Fig.~\ref{FigSymCosetRes}. This diagram corresponds to the case $\mathfrak{g} = V_{0} \oplus V_{1}$, with $V_{0}$ a subalgebra and $V_{1}$ a symmetric coset, and the choice $S = S_{\mathrm{E}}^{\left( 3 \right)}$.

The subspaces of $\mathfrak{g}$ are represented on the horizontal axis, while the semigroup elements occupy the vertical one. In this way, the whole $S_{\mathrm{E}}^{\left( 3 \right)}$-expanded algebra $S_{\mathrm{E}}^{\left( 3 \right)} \times \mathfrak{g}$ corresponds to the shaded region in Fig.~\ref{FigSymCosetRes}~(\textit{a}). In Fig.~\ref{FigSymCosetRes}~(\textit{b}), the gray region represents the resonant subalgebra $\mathfrak{G}_{\mathrm{R}} = W_{0} \oplus W_{1}$ with
\begin{eqnarray}
S_{0} & = & \left\{ \lambda_{0}, \lambda_{2}, \lambda_{4} \right\}, \label{partX1} \\
S_{1} & = & \left\{ \lambda_{1}, \lambda_{3}, \lambda_{4} \right\}. \label{partX2}
\end{eqnarray}

Let us observe that each column in the diagram corresponds to a subset of the resonant partition. Finally, Fig.~\ref{FigSymCosetRes}~(\textit{c}) represents the $0_{S}$-reduced algebra, obtained after imposing $\lambda_{4} \times \mathfrak{g} = 0$. This figure actually corresponds to the case $\mathcal{G} \left( N_{0}, N_{1} \right)$.

As is evident from the above discussion, the case $N=1$, $\lambda
_{N+1}\bm{T}_{A}=\bm{0}$ reproduces the
\.{I}n\"{o}n\"{u}--Wigner contraction for $\mathfrak{g} = V_{0} \oplus V_{1}$. More on generalized \.{I}n\"{o}n\"{u}--Wigner contractions is
presented in section~\ref{SecOther}.

\subsubsection{\label{Sec W-W ResSubAlg}Case when $\mathfrak{g}$ fulfills the Weimar-Woods Conditions}

Let $\mathfrak{g} = \bigoplus_{p=0}^{n} V_{p}$ be a subspace decomposition of $\mathfrak{g}$. In terms of this decomposition, the Weimar-Woods conditions~\cite{W-W,W-W2} on $\mathfrak{g}$ read
\begin{equation}
\left[ V_{p}, V_{q} \right] \subset \bigoplus_{r=0}^{H_{n} \left( p+q \right)} V_{r}. \label{Weimar-Woods Condition}
\end{equation}

Let
\begin{equation}
S_{\mathrm{E}} = \bigcup_{p=0}^{n} S_{p}
\end{equation}
be a subset decomposition of $S_{\mathrm{E}}$, where the subsets
$S_{p}\subset S_{\mathrm{E}}$ are defined by%
\begin{equation}
S_{p}=\left\{  \lambda_{\alpha_{p}}\text{, such that }\alpha_{p}%
=p,\ldots,N+1\right\}
\end{equation}
with $N+1\geq n$.

This subset decomposition is a resonant one under the semigroup
product~(\ref{Normal Expansion Product}), because it satisfies [compare
eq.~(\ref{Sp x Sq = SHn(p+q)}) with eq.~(\ref{Weimar-Woods Condition})]
\begin{equation}
S_{p} \cdot S_{q} = S_{H_{n} \left( p+q \right)} \subset \bigcap_{r=0}^{H_{n} \left( p+q \right)} S_{r}.
\label{Sp x Sq = SHn(p+q)}
\end{equation}

According to Theorem~\ref{ResSubAlgTh}, the direct sum
\begin{equation}
\mathfrak{G}_{\mathrm{R}} = \bigoplus_{p=0}^{n} W_{p},
\end{equation}
with
\[
W_{p} = S_{p} \times V_{p},
\]
is a resonant subalgebra of $\mathfrak{G}$.

Using eq.~(\ref{ResonantStructureConstants}), we get the following structure
constants for the resonant subalgebra:%
\[
C_{\left( a_{p}, \alpha_{p} \right) \left( b_{q}, \beta_{q} \right)}^{\phantom{\left( a_{p}, \alpha_{p} \right) \left( b_{q}, \beta_{q} \right)} \left( c_{r}, \gamma_{r} \right)} = \delta_{H_{N+1} \left( \alpha_{p} + \beta_{q} \right)}^{\gamma_{r}} C_{a_{p} b_{q}}^{\phantom{a_{p} b_{q}} c_{r}} \text{ with }\left\{
\begin{array}
[c]{l}%
p,q,r=0,\ldots,n\\
\alpha_{p},\beta_{p},\gamma_{p}=p,\ldots,N+1
\end{array}
\right.  .
\]
Imposing $\lambda_{N+1}\bm{T}_{A}=\bm{0}$, this becomes
\begin{equation}
C_{\left( a_{p}, \alpha_{p} \right) \left( b_{q}, \beta_{q} \right)}^{\phantom{\left( a_{p}, \alpha_{p} \right) \left( b_{q}, \beta_{q} \right)} \left( c_{r}, \gamma_{r} \right)} = \delta_{H_{N+1} \left( \alpha_{p} + \beta_{q} \right)}^{\gamma_{r}} C_{a_{p} b_{q}}^{\phantom{a_{p} b_{q}} c_{r}} \text{ with }\left\{
\begin{array}
[c]{l}%
p,q,r=0,\ldots,n\\
\alpha_{p},\beta_{p},\gamma_{p}=p,\ldots,N
\end{array}
\right. .
\label{xxxStrucConstZampForz}
\end{equation}

This $0_{S}$-reduced algebra corresponds to the case $\mathcal{G}\left(
N_{0},\ldots,N_{n}\right)  $ of Theorem 3 from~\cite{Azcarraga Et Al} with
$N_{p}=N$ for every $p=0,\ldots,n$.  The structure constants~(\ref{xxxStrucConstZampForz}) correspond to the ones of eq.~(4.8) in Ref.~\cite{Azcarraga Et Al} (with a slightly different notation). The more general case,
\[
N_{p+1}=\left\{
\begin{array}
[c]{l}%
N_{p}\text{ or}\\
N_{p}+1
\end{array}
\right.
\]
can be also obtained in the context of an $S$-expansion, from a resonant subalgebra and applying a kind of reduction more general than the  $0_{S}$-reduction (see sec.~\ref{SecOther} and app.~\ref{appX}).

\begin{figure}
\includegraphics[width=\columnwidth]{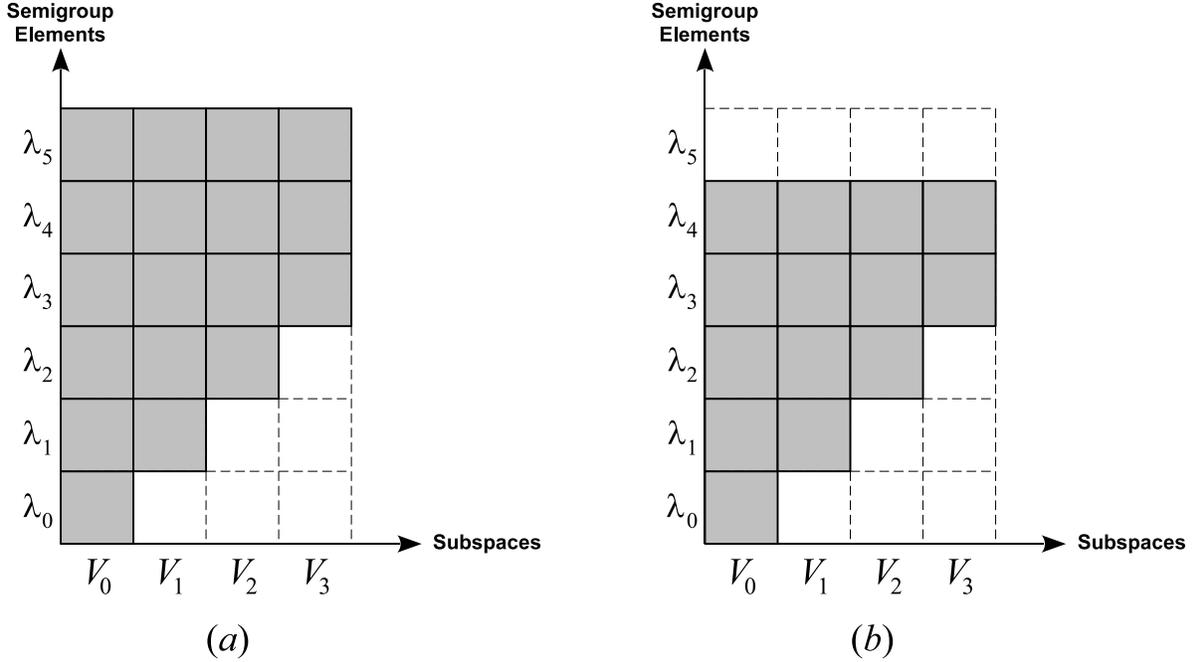}
\caption{\label{FigW-WResonantSubAlg}(\textit{a}) The shaded region shows a $S_{\mathrm{E}}^{(4)}$ resonant subalgebra when $\mathfrak{g} = V_{0} \oplus V_{1} \oplus V_{2} \oplus V_{3}$ satisfies the Weimar-Woods conditions. (\textit{b}) The $0_{S}$-reduction of this resonant subalgebra removes all sectors of the form $0_{S} \times \mathfrak{g}$. This corresponds to the case $\mathcal{G}(4,4,4,4)$ in the context of~\cite{Azcarraga Et Al}.}
\end{figure}

The resonant subalgebra for the Weimar-Woods case with $n=3$, $N=4$ and its $0_{S}$-reduction are shown in Fig.~\ref{FigW-WResonantSubAlg}~(\textit{a}) and~(\textit{b}), respectively.

\subsubsection{Case when $\mathfrak{g}=V_{0}\oplus V_{1}\oplus V_{2}$ is a
Superalgebra}

A superalgebra $\mathfrak{g}$ comes naturally split into three subspaces
$V_{0}$, $V_{1}$ and $V_{2}$, where $V_{1}$ corresponds to the Fermionic
sector and $V_{0}\oplus V_{2}$ to the Bosonic one, with
$V_{0}$ being a subalgebra. This subspace structure may be written as
\begin{eqnarray}
\left[  V_{0},V_{0}\right]   &  \subset & V_{0},\label{SuperV0V0=V0}\\
\left[  V_{0},V_{1}\right]   &  \subset & V_{1},\label{SuperV0V1=V1}\\
\left[  V_{0},V_{2}\right]   &  \subset & V_{2},\label{SuperV0V2=V2}\\
\left[  V_{1},V_{1}\right]   &  \subset & V_{0}\oplus V_{2}%
,\label{SuperV1V1=V0+V2}\\
\left[  V_{1},V_{2}\right]   &  \subset & V_{1},\label{SuperV1V2=V1}\\
\left[  V_{2},V_{2}\right]   &  \subset & V_{0}\oplus V_{2}%
.\label{SuperV2V2=V0+V2}%
\end{eqnarray}

Let $S_{\mathrm{E}}^{(N)}=S_{0}\cup S_{1}\cup S_{2}$ be a subset
decomposition of $S_{\mathrm{E}}^{(N)}$, where the subsets $S_{0},S_{1}%
,S_{2}$ are given by the general expression%
\begin{equation}
S_{p}=\left\{  \lambda_{2m+p},\text{ with }m=0,\ldots,\left[  \frac{N-p}{2} \right]  \right\}  \cup\left\{  \lambda_{N+1}\right\}  ,\qquad p=0,1,2.
\label{ResSuperPartition Sp}%
\end{equation}

This subset decomposition is a resonant one, because it satisfies [compare
eqs.~(\ref{SuperS0S0=S0})--(\ref{SuperS2S2=S0nS2}) with eqs.~(\ref{SuperV0V0=V0})--(\ref{SuperV2V2=V0+V2})]
\begin{eqnarray}
S_{0}\cdot S_{0} &  \subset & S_{0},\label{SuperS0S0=S0}\\
S_{0}\cdot S_{1} &  \subset & S_{1},\label{SuperS0S1=S1}\\
S_{0}\cdot S_{2} &  \subset & S_{2},\label{SuperS0S2=S2}\\
S_{1}\cdot S_{1} &  \subset & S_{0}\cap S_{2},\label{SuperS1S1=S0nS2}\\
S_{1}\cdot S_{2} &  \subset & S_{1},\label{SuperS1S2=S1}\\
S_{2}\cdot S_{2} &  \subset & S_{0}\cap S_{2}.\label{SuperS2S2=S0nS2}%
\end{eqnarray}

Theorem~\ref{ResSubAlgTh} assures us that $\mathfrak{G}_{\mathrm{R}} = W_{0} \oplus W_{1} \oplus W_{2}$, with $W_{p} = S_{p} \times V_{p}$, $p=0,1,2$, is a resonant subalgebra.

Using eq.~(\ref{ResonantStructureConstants}), it is possible to write down the structure constants for the resonant subalgebra as
\begin{equation}
C_{\left( a_{p}, \alpha_{p} \right) \left( b_{q}, \beta_{q} \right)}^{\phantom{\left( a_{p}, \alpha_{p} \right) \left( b_{q}, \beta_{q} \right)} \left( c_{r}, \gamma_{r} \right)} = \delta_{H_{N+1} \left( \alpha_{p} + \beta_{q} \right)}^{\gamma_{r}} C_{a_{p} b_{q}}^{\phantom{a_{p} b_{q}} c_{r}} \text{ with }
\left\{
\begin{array}[c]{l}
p,q,r=0,1,2, \\
\alpha_{p},\beta_{p},\gamma_{p}=2m+p, \\
m=0,\ldots,\left[  \frac{N-p}{2}\right] ,\frac{N+1-p}{2}.
\end{array}
\right.
\end{equation}

Imposing $\lambda_{N+1}\bm{T}_{A}=\bm{0}$, the structure
constants for the $0_{S}$-reduction of the resonant subalgebra are obtained:
\begin{equation}
C_{\left( a_{p}, \alpha_{p} \right) \left( b_{q}, \beta_{q} \right)}^{\phantom{\left( a_{p}, \alpha_{p} \right) \left( b_{q}, \beta_{q} \right)} \left( c_{r}, \gamma_{r} \right)} = \delta_{H_{N+1} \left( \alpha_{p} + \beta_{q} \right)}^{\gamma_{r}} C_{a_{p} b_{q}}^{\phantom{a_{p} b_{q}} c_{r}} \text{ with }
\left\{
\begin{array}[c]{l}
p,q,r=0,1,2, \\
\alpha_{p},\beta_{p},\gamma_{p}=2m+p, \\
m=0,\ldots,\left[  \frac{N-p}{2}\right] .
\end{array}
\right.
\label{SuperResonantStructureConstants}%
\end{equation}

This $0_{S}$-reduced algebra corresponds to the algebra $\mathcal{G} \left( N_{0}, N_{1}, N_{2} \right)$ with
\[
N_{p}=2\left[  \frac{N-p}{2}\right]  +p,\qquad p=0,1,2
\]
found in theorem~5 of Ref.~\cite{Azcarraga Et Al}. The structure constants~(\ref{SuperResonantStructureConstants}) match the structure constants~(5.6) from Ref.~\cite{Azcarraga Et Al}.

\begin{figure}
\includegraphics[width=\columnwidth]{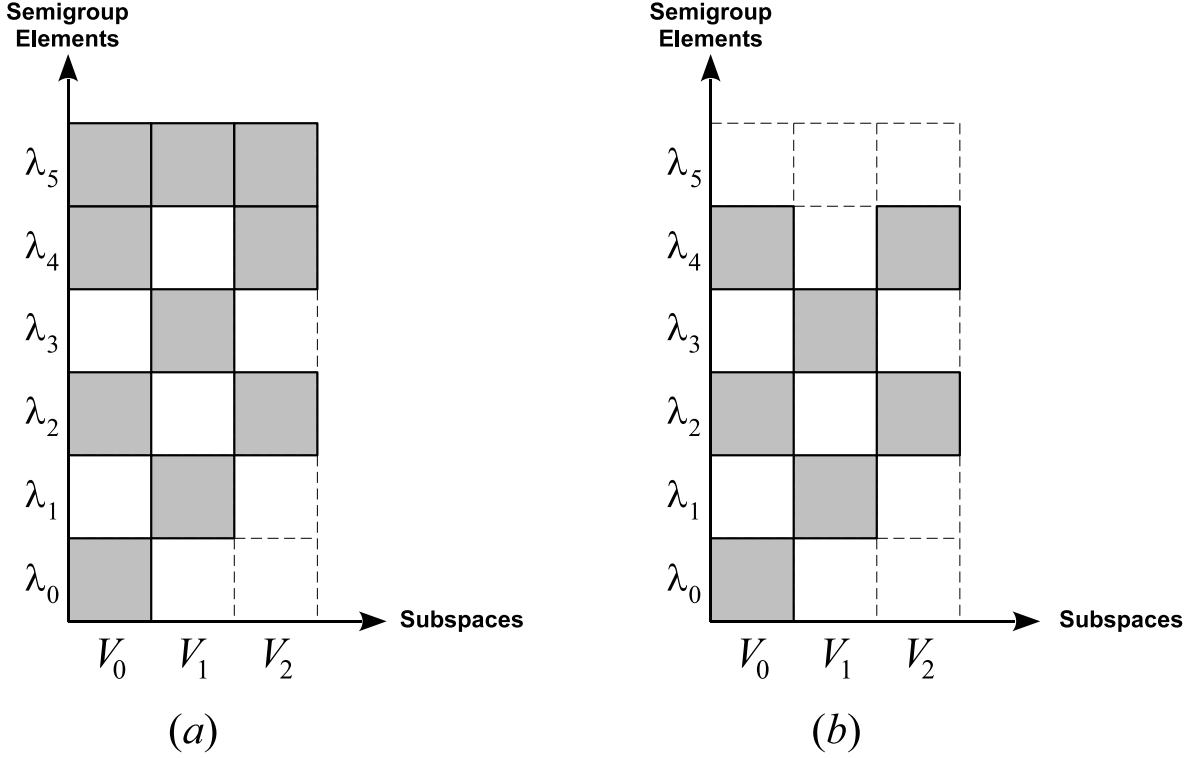}
\caption{\label{FigSupResSubAlg}(\textit{a}) The shaded area corresponds to an $S_{\mathrm{E}}^{(4)}$ resonant subalgebra of $\mathfrak{G} = S_{\mathrm{E}}^{(4)} \times \mathfrak{g}$ when $\mathfrak{g}$ is a superalgebra. (\textit{b}) The gray region shows the $0_{S}$-reduction of the resonant subalgebra $\mathfrak{G}_{\mathrm{R}}$. This corresponds to $\mathcal{G}(4,3,4)$ in the context of~\cite{Azcarraga Et Al}.}
\end{figure}

Fig.~\ref{FigSupResSubAlg}~(\textit{a}) shows the resonant subalgebra for the case of superalgebras, and Fig.~\ref{FigSupResSubAlg}~(\textit{b}), its corresponding $0_{S}$-reduction, for the case $N=4$.

\section{\label{SecSExpExamples}$S$-Expansions of $\mathfrak{osp}\left(
32|1\right)  $ and $d=11$ Superalgebras}

In this section we explore some explicit examples of $S$-expansions. In
general, every possible choice of abelian semigroup $S$ and resonant partition will lead to a new $d=11$ superalgebra. Note however that the
existence of a resonant partition is not at all guaranteed for an arbitrary
semigroup $S$.

Since our main physical motivation comes from 11-dimensional Supergravity, we shall always take the orthosymplectic superalgebra $\mathfrak{osp} \left( 32|1 \right)$ as a starting point. A suitable basis is provided by $\left\{ \bm{P}_{a}, \bm{J}_{ab}, \bm{Z}_{abcde}, \bm{Q} \right\}$, where $\left\{ \bm{P}_{a}, \bm{J}_{ab} \right\}$ are the anti-de Sitter (AdS) generators, $\bm{Z}_{abcde}$ is a 5-index antisymmetric tensor and $\bm{Q}$ is a 32-component, Majorana spinor charge. The $\mathfrak{osp} \left( 32|1 \right)$ (anti)commutation relations explicitly read
\begin{eqnarray}
\left[ \bm{P}_{a}, \bm{P}_{b}\right] & = & \bm{J}_{ab},\\
\left[ \bm{J}^{ab}, \bm{P}_{c} \right] & = & \delta_{ec}^{ab} \bm{P}^{e},\\
\left[ \bm{J}^{ab}, \bm{J}_{cd} \right] & = & \delta_{ecd}^{abf} \bm{J}_{\phantom{e} f}^{e},
\end{eqnarray}
\begin{eqnarray}
\left[ \bm{P}_{a}, \bm{Z}_{b_{1} \cdots b_{5}} \right] & = & - \frac{1}{5!} \varepsilon_{ab_{1} \cdots b_{5} c_{1} \cdots c_{5}} \bm{Z}^{c_{1} \cdots c_{5}},\\
\left[ \bm{J}^{ab}, \bm{Z}_{c_{1}\cdots c_{5}} \right] & = & \frac{1}{4!}\delta_{dc_{1} \cdots c_{5}}^{abe_{1} \cdots e_{4}} \bm{Z}_{\phantom{d} e_{1} \cdots e_{4}}^{d},
\end{eqnarray}%
\begin{eqnarray}
\left[  \bm{Z}^{a_{1}\cdots a_{5}},\bm{Z}_{b_{1}\cdots b_{5}%
}\right]   &  = & \eta^{\left[  a_{1}\cdots a_{5}\right]  \left[  c_{1}\cdots
c_{5}\right]  }\varepsilon_{c_{1}\cdots c_{5}b_{1}\cdots b_{5}e}%
\bm{P}^{e}+\delta_{db_{1}\cdots b_{5}}^{a_{1}\cdots a_{5}%
e}\bm{J}_{\phantom{d} e}^{d}+\nonumber\\
&  & -\frac{1}{3!3!5!}\varepsilon_{c_{1}\cdots c_{11}}\delta_{d_{1}d_{2}%
d_{3}b_{1}\cdots b_{5}}^{a_{1}\cdots a_{5}c_{4}c_{5}c_{6}}\eta^{\left[
c_{1}c_{2}c_{3}\right]  \left[  d_{1}d_{2}d_{3}\right]  }\bm{Z}%
^{c_{7}\cdots c_{11}},
\end{eqnarray}
\begin{eqnarray}
\left[  \bm{P}_{a},\bm{Q}\right]   &  = & -\frac{1}{2}\Gamma
_{a}\bm{Q},\\
\left[  \bm{J}_{ab},\bm{Q}\right]   &  = & -\frac{1}{2}%
\Gamma_{ab}\bm{Q},\\
\left[  \bm{Z}_{abcde},\bm{Q}\right]   &  = & -\frac{1}{2}%
\Gamma_{abcde}\bm{Q},
\end{eqnarray}%
\begin{eqnarray}
\left\{  \bm{Q}^{\rho},\bm{Q}^{\sigma}\right\}   &  = & -\frac
{1}{2^{3}}\left[  \left(  \Gamma^{a}C^{-1}\right)  ^{\rho\sigma}%
\bm{P}_{a}-\frac{1}{2}\left(  \Gamma^{ab}C^{-1}\right)  ^{\rho\sigma
}\bm{J}_{ab}+\right.  \nonumber\\
& & \left.  +\frac{1}{5!}\left(  \Gamma^{abcde}C^{-1}\right)  ^{\rho\sigma
}\bm{Z}_{abcde}\right]  ,
\end{eqnarray}
where $C_{\rho\sigma}$ is the charge conjugation matrix and $\Gamma_{a}$ are Dirac matrices in 11 dimensions.

As a first step towards the $S$-Expansion, the $\mathfrak{osp}\left(
32|1\right)  $ algebra is written as the direct sum of three subspaces:
\begin{equation}
\mathfrak{osp}\left(  32|1\right)  =V_{0}\oplus V_{1}\oplus V_{2},
\label{ospSubSep}%
\end{equation}
where $V_{0}$ corresponds to the Lorentz subalgebra (spanned by
$\bm{J}_{ab}$), $V_{1}$ to the Fermionic subspace (spanned by
$\bm{Q}$) and $V_{2}$ to the translations and the M5-brane piece
(spanned by $\bm{P}_{a}$ and $\bm{Z}_{abcde}$). The subspace
separation (\ref{ospSubSep}) satisfies conditions~(\ref{SuperV0V0=V0}%
)--(\ref{SuperV2V2=V0+V2}), as can be easily checked.

The M~algebra~\cite{Towsend1,Towsend2,CECS M} and a Superalgebra similar to those of D'Auria--Fr\'{e}~\cite{DAuria-Fre} are rederived in next
sections using $S=S_{\mathrm{E}}^{\left(  N\right)  }$ with $N=2$ and
$N=3$, respectively. As an example of an $S$-expansion with $S\neq
S_{\mathrm{E}}$, the case $S=\mathbb{Z}_{4}$ is considered in
section~\ref{SecZ4Exp}, where a new superalgebra resembling aspects of
the M~algebra, $\mathfrak{osp}\left(  32|1\right)  \times\mathfrak{osp}%
\left(  32|1\right)  $ and D'Auria--Fr\'{e} superalgebras is found.

The inclusion of the M~algebra among the examples draws from two different but related facts. First, the $S$-Expansion paradigm casts the M~algebra as one from a family of superalgebras, all derived from $\mathfrak{osp}\left(  32|1\right)$ through different choices for the semigroup and different alternatives of reduction, when present at all. This can be relevant from a physical point of view, since all of them share important features. The second reason deals with the construction of Chern--Simons and transgression Lagrangians. As will be shown in sec.~\ref{SecInvTen}, invariant tensors for resonant subalgebras and $0_{S}$-reduced algebras thereof are readily available, but this is not the case for general reduced algebras. As such, the fact that the M~algebra stems from a $0_{S}$-reduction is interesting not only because it provides with an invariant tensor derived from $\mathfrak{osp}\left(  32|1\right)$, but also because it brings about the possibility of considering its direct generalization, namely, the resonant subalgebra from where it was extracted. More on the physical consequences of regarding the M~algebra as the $0_{S}$-reduction of a resonant subalgebra is found in~\cite{Iza06c}.

\subsection{\label{SecMAlgResSub}The M~Algebra}

As treated in detail in~\cite{Azcarraga Et Al,AzcarragaSuperspace}, the complete M~algebra (i.e., including its Lorentz part) can be obtained by means of an MC expansion of $\mathfrak{osp}\left(  32|1\right)  $. Within the present scheme, the M~algebra should be recovered via an $S$-expansion with $S = S_{\mathrm{E}}^{\left( 2 \right)}$ followed by a $0_{S}$-reduction, as explained in section~\ref{SecAlgExpSExp}.

\begin{figure}
\includegraphics[width=\columnwidth]{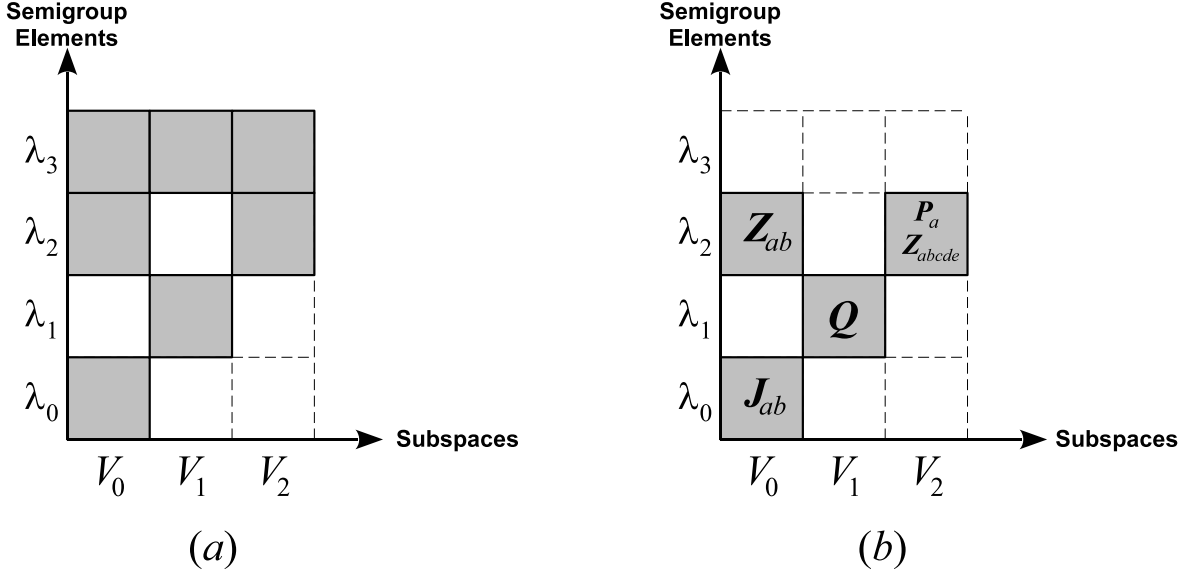}
\caption{\label{Fig MAlgebra}The M~algebra as an $S_{\mathrm{E}}^{\left( 2 \right)}$-expansion of $\mathfrak{osp} \left( 32|1 \right)$. (\textit{a}) A resonant subalgebra of the $S_{\mathrm{E}}^{\left( 2 \right)}$-expanded algebra $\mathfrak{G} = S_{\mathrm{E}}^{\left( 2 \right)} \times \mathfrak{osp} \left( 32|1 \right)$ is shown in the shaded region. (\textit{b}) The M~algebra itself (gray area) is obtained after $0_{S}$-reducing the resonant subalgebra.}
\end{figure}

In order to obtain the M~algebra in the context of $S$-expansions, one has to
pick $S_{\mathrm{E}}^{\left(  2\right)  }=\left\{  \lambda_{0}%
,\lambda_{1},\lambda_{2},\lambda_{3}\right\}  $, use the resonant
partition~(\ref{ResSuperPartition Sp}) and impose the condition $\lambda
_{3}\bm{T}_{A}=\bm{0}$. This amounts to explicitly writing the
structure constants~(\ref{SuperResonantStructureConstants}) for this case. For
the sake of simplicity, let us relabel $\bm{J}_{ab}=\bm{J}%
_{\left(  ab,0\right)  }$, $\bm{Q}_{\alpha}=\bm{Q}_{\left(
\alpha,1\right)  }$, $\bm{Z}_{ab}=\bm{J}_{\left(  ab,2\right)
}$, $\bm{P}_{a}=\bm{P}_{\left(  a,2\right)  }$,
$\bm{Z}_{abcde}=\bm{Z}_{\left(  abcde,2\right)  }$, as shown
in Fig.~\ref{Fig MAlgebra}. The resulting algebra reads
\begin{eqnarray}
\left[  \bm{P}_{a},\bm{P}_{b}\right]   &  = & \bm{0},\\
\left[ \bm{J}^{ab}, \bm{P}_{c} \right] & = & \delta_{ec}^{ab} \bm{P}^{e},\\
\left[ \bm{J}^{ab}, \bm{J}_{cd} \right] & = & \delta_{ecd}^{abf} \bm{J}_{\phantom{e} f}^{e},
\end{eqnarray}%
\begin{eqnarray}
\left[  \bm{J}^{ab},\bm{Z}_{cd}\right]   &  = & \delta_{ecd}^{abf} \bm{Z}_{\phantom{e} f}^{e},\\
\left[  \bm{Z}^{ab},\bm{Z}_{cd}\right] & = & \bm{0},\\
\left[  \bm{P}_{a},\bm{Z}_{b_{1}\cdots b_{5}}\right]   &
= & \bm{0},\\
\left[ \bm{J}^{ab}, \bm{Z}_{c_{1}\cdots c_{5}} \right] & = & \frac{1}{4!}\delta_{dc_{1} \cdots c_{5}}^{abe_{1} \cdots e_{4}} \bm{Z}_{\phantom{d} e_{1} \cdots e_{4}}^{d},\\
\left[  \bm{Z}^{ab},\bm{Z}_{c_{1}\cdots c_{5}}\right]
&  = & \bm{0},\\
\left[  \bm{Z}^{a_{1}\cdots a_{5}},\bm{Z}_{b_{1}\cdots b_{5}%
}\right]   &  = & \bm{0},
\end{eqnarray}%
\begin{eqnarray}
\left[  \bm{P}_{a},\bm{Q}\right]   &  = & \bm{0},\\
\left[  \bm{J}_{ab},\bm{Q}\right]   &  = & -\frac{1}{2}%
\Gamma_{ab}\bm{Q},\\
\left[  \bm{Z}_{ab},\bm{Q}\right]   &  = & \bm{0},\\
\left[  \bm{Z}_{abcde},\bm{Q}\right]   &  = & \bm{0},
\end{eqnarray}%
\begin{eqnarray}
\left\{  \bm{Q}^{\rho},\bm{Q}^{\sigma}\right\}   &  = & -\frac
{1}{2^{3}}\left[  \left(  \Gamma^{a}C^{-1}\right)  ^{\rho\sigma}%
\bm{P}_{a}-\frac{1}{2}\left(  \Gamma^{ab}C^{-1}\right)  ^{\rho\sigma
}\bm{Z}_{ab}+\right. \nonumber\\
&  & \left.  +\frac{1}{5!}\left(  \Gamma^{abcde}C^{-1}\right)  ^{\rho\sigma
}\bm{Z}_{abcde}\right]  .
\end{eqnarray}

Note that the r\^{o}le of the $0_{S}$-reduction in the process is that of
abelianizing large sectors of the resonant subalgebra.

\subsection{D'Auria--Fr\'{e}-like Superalgebra}

The above example used $S_{\mathrm{E}}^{\left( 2 \right)} = \left\{ \lambda_{0}, \lambda_{1}, \lambda_{2}, \lambda_{3} \right\}$ as abelian semigroup to perform the expansion. In this section, the results of choosing instead $S_{\mathrm{E}}^{\left( 3 \right)} = \left\{ \lambda_{0}, \lambda_{1}, \lambda_{2}, \lambda_{3}, \lambda_{4} \right\}$ while leaving
everything else (including the $0_{S}$-reduction) unchanged are examined.

\begin{figure}
\includegraphics[width=\columnwidth]{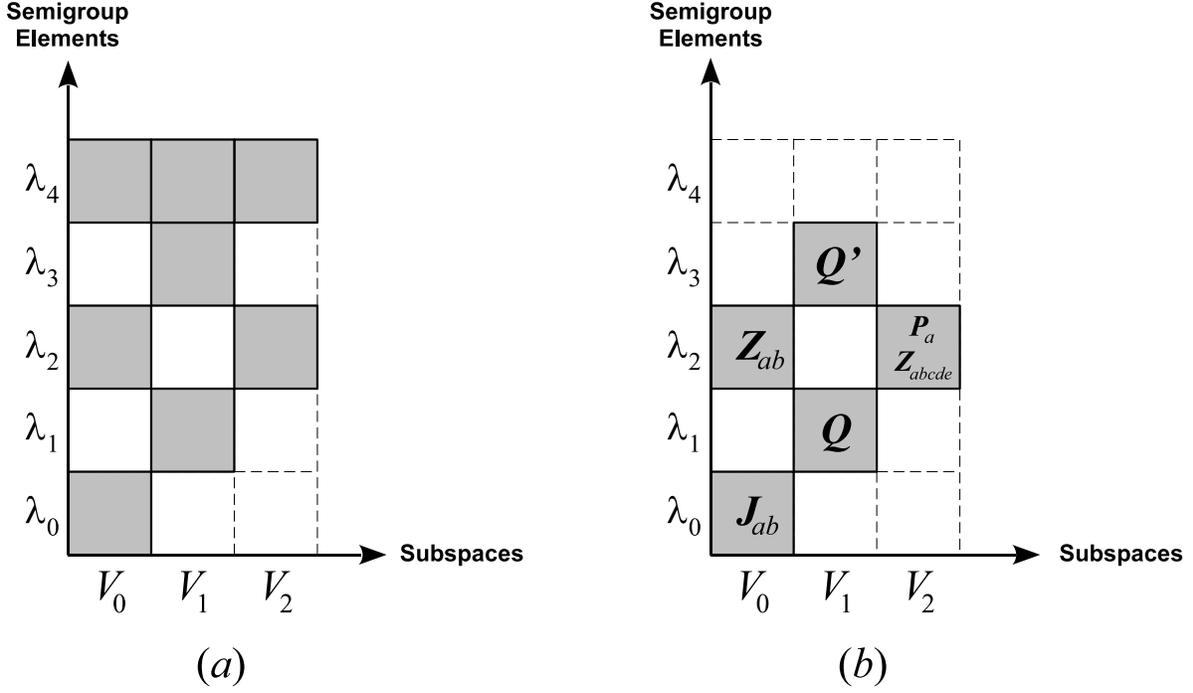}
\caption{\label{Fig Res DAuriaFre}A D'Auria--Fr\'{e}-like Superalgebra regarded here as an $S_{\mathrm{E}}^{\left( 3 \right)}$-expansion of $\mathfrak{osp} \left( 32|1 \right)$. (\textit{a}) A resonant subalgebra of the $S_{\mathrm{E}}^{\left( 3\right)}$-expanded algebra $\mathfrak{G} = S_{\mathrm{E}}^{\left( 3 \right)} \times \mathfrak{osp} \left( 32|1 \right)$ is shown in the shaded region. (\textit{b}) A Superalgebra similar to the ones introduced by D'Auria and Fr\'{e} in~\cite{DAuria-Fre} is obtained after $0_{S}$-reducing the resonant subalgebra.}
\end{figure}

A D'Auria--Fr\'{e}-like Superalgebra~\cite{DAuria-Fre}, with one extra
Fermionic generator as compared with $\mathfrak{osp} \left( 32|1 \right)$ or the M~algebra, is obtained by picking the resonant partition~(\ref{ResSuperPartition Sp}) and $0_{S}$-reducing the resulting resonant subalgebra. Relabeling generators as $\bm{J}_{ab}%
=\bm{J}_{\left(  ab,0\right)  }$, $\bm{Q}_{\alpha
}=\bm{Q}_{\left(  \alpha,1\right)  }$, $\bm{Z}_{ab}%
=\bm{J}_{\left(  ab,2\right)  }$, $\bm{P}_{a}=\bm{P}%
_{\left(  a,2\right)  }$, $\bm{Z}_{abcde}=\bm{Z}_{\left(
abcde,2\right)  }$, $\bm{Q}_{\alpha}^{\prime}=\bm{Q}_{\left(
\alpha,3\right)  }$, one finds the structure depicted in
Fig.~\ref{Fig Res DAuriaFre}. While this algebra has the same \emph{structure} (i.e., same number and type of generators, with commutators valued on the same subspaces) as the ones introduced by D'Auria and Fr\'{e} in~\cite{DAuria-Fre}, the details differ, so it cannot really correspond to any of them~\footnote{\textit{Note added}: The algebra here considered and the original ones from D'Auria--Fr\'{e} correspond to different members of a family of superalgebras introduced in Ref.~\cite{Ban04}.} (hence the ``-like'').

The (anti)commutation relations, which can be read off directly from the
structure constants~(\ref{SuperResonantStructureConstants}) after applying
condition~(\ref{0sxT=0}), bear a strong similarity with those from the
M~algebra. Only the following three differ:
\begin{eqnarray}
\left[  \bm{P}_{a},\bm{Q}\right]   &  = & -\frac{1}{2}\Gamma
_{a}\bm{Q}^{\prime},\\
\left[  \bm{Z}_{ab},\bm{Q}\right]   &  = & -\frac{1}{2}%
\Gamma_{ab}\bm{Q}^{\prime},\\
\left[  \bm{Z}_{abcde},\bm{Q}\right]   &  = & -\frac{1}{2}%
\Gamma_{abcde}\bm{Q}^{\prime}.
\end{eqnarray}
The (anti)commutation relations which directly involve the extra Fermionic
generator $\bm{Q}^{\prime}$ read
\begin{eqnarray}
\left[  \bm{P}_{a},\bm{Q}^{\prime}\right]   &  = & \bm{0}%
,\label{pqp}\\
\left[  \bm{Z}_{ab},\bm{Q}^{\prime}\right]   &
= & \bm{0},\label{z2qp}\\
\left[  \bm{Z}_{abcde},\bm{Q}^{\prime}\right]   &
= & \bm{0},\label{z5qp}\\
\left\{  \bm{Q},\bm{Q}^{\prime}\right\}   &  = & \bm{0}%
,\\
\left\{  \bm{Q}^{\prime},\bm{Q}^{\prime}\right\}   &
= & \bm{0},
\end{eqnarray}%
\begin{equation}
\left[  \bm{J}_{ab},\bm{Q}^{\prime}\right]  =-\frac{1}%
{2}\Gamma_{ab}\bm{Q}^{\prime}.
\end{equation}
The extra Fermionic generator $\bm{Q}^{\prime}$ is found to
(anti)commute with all generators from the algebra but the Lorentz generators
(which was to be expected due to its spinor character).

\subsection{\label{SecZ4Exp}Resonant Subalgebra of $\mathbb{Z}_{4}%
\times\mathfrak{osp}\left(  32|1\right)  $}

Cyclic groups seem especially suitable for an $S$-expansion, because, on one hand, they are groups and not only semigroups (and therefore there is no $0_{S}$ element and no $0_{S}$-reduction), and on the other, because the multiplication law for a cyclic group looks very similar to the multiplication law of the semigroup $S_{\mathrm{E}}$,
\begin{eqnarray}
S_{\mathrm{E}}^{\left(  N\right)  }  &  : & \lambda_{\alpha}\lambda_{\beta
}=\lambda_{H_{N+1}\left(  \alpha+\beta\right)  }\\
\mathbb{Z}_{N}  &  : & \lambda_{\alpha}\lambda_{\beta}=\lambda
_{\operatorname{mod}_{N}\left(  \alpha+\beta\right)  }.
\end{eqnarray}

The cyclic group $\mathbb{Z}_{4}$ in particular was chosen for this example because the $\mathbb{Z}_{2}$ case is trivial [the resonant subalgebra is $\mathfrak{osp} \left( 32|1 \right)$ itself] and $\mathbb{Z}_{3}$ seems to have no resonant partition; therefore $\mathbb{Z}_{4}$ corresponds to the simplest nontrivial case.

Since this example uses a semigroup different from $S_{\mathrm{E}}$, the algebra obtained does not correspond to a Maurer--Cartan forms power-series expansion.

Given a superalgebra $\mathfrak{g}=V_{0}\oplus V_{1}\oplus V_{2}$ with the
structure~(\ref{SuperV0V0=V0})--(\ref{SuperV2V2=V0+V2}), a resonant partition
of $\mathbb{Z}_{4}=\left\{  \lambda_{0},\lambda_{1},\lambda_{2},\lambda
_{3}\right\}  $ is given by
\begin{eqnarray}
S_{0} & = & \left\{ \lambda_{0}, \lambda_{2} \right\} , \\
S_{1} & = & \left\{ \lambda_{1}, \lambda_{3} \right\} , \\
S_{2} & = & \left\{ \lambda_{0}, \lambda_{2} \right\} .
\end{eqnarray}

\begin{figure}
\includegraphics[width=.5\columnwidth]{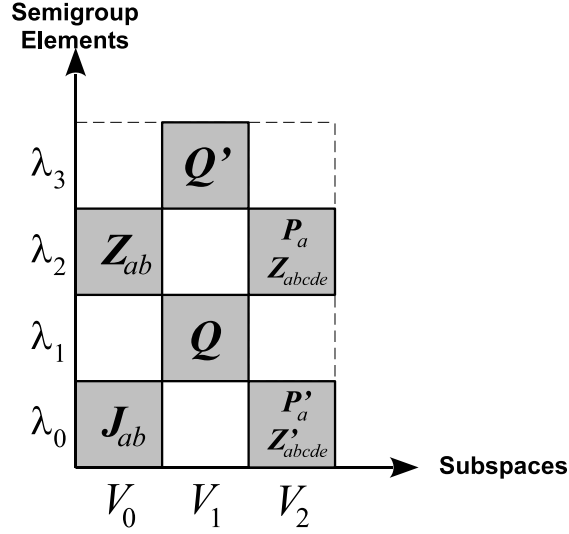}
\caption{\label{Fig ResonantZ4}A new $d=11$ Superalgebra, different from but resembling aspects of both the M~algebra and the D'Auria--Fr\'{e} Superalgebras, is obtained directly as a resonant subalgebra (shaded region) of the $S$-expanded algebra $\mathbb{Z}_{4} \times\mathfrak{osp} \left(32|1\right)$.}
\end{figure}

In order to avoid a cluttering of indices, relabel $\bm{J}%
_{ab}=\bm{J}_{\left(  ab,0\right)  }$, $\bm{Z}_{a_{1}\cdots
a_{5}}^{\prime}=\bm{Z}_{\left(  a_{1}\cdots a_{5},0\right)  }$,
$\bm{P}_{a}^{\prime}=\bm{P}_{\left(  a,0\right)  }$,
$\bm{Q}_{\alpha}=\bm{Q}_{\left(  \alpha,1\right)  }$,
$\bm{Z}_{ab}=\bm{J}_{\left(  ab,2\right)  }$, $\bm{Z}%
_{a_{1}\cdots a_{5}}=\bm{Z}_{\left(  a_{1}\cdots a_{5},2\right)  }$,
$\bm{P}_{a}=\bm{P}_{\left(  a,2\right)  }$ and $\bm{Q}%
_{\alpha}^{\prime}=\bm{Q}_{\left(  \alpha,3\right)  }$, as shown in
Fig.~\ref{Fig ResonantZ4}.

The full resonant subalgebra can be easily obtained using the structure
constants~(\ref{ResonantStructureConstants}); here we shall only quote some of its more interesting sectors.

This algebra has two very conspicuous features: first, it bears \emph{some}
resemblance to the M~algebra, for \emph{both} Fermionic generators
$\bm{Q}$ and $\bm{Q}^{\prime}$ satisfy
\begin{eqnarray}
\left\{  \bm{Q}^{\prime\rho},\bm{Q}^{\prime\sigma}\right\}
&  = & \left\{  \bm{Q}^{\rho},\bm{Q}^{\sigma}\right\}
\nonumber\\
&  = & -\frac{1}{2^{3}}\left[  \left(  \Gamma^{a}C^{-1}\right)  ^{\rho\sigma
}\bm{P}_{a}-\frac{1}{2}\left(  \Gamma^{ab}C^{-1}\right)  ^{\rho\sigma
}\bm{Z}_{ab}+\right.  \nonumber\\
&  &  \left.  +\frac{1}{5!}\left(  \Gamma^{abcde}C^{-1}\right)  ^{\rho\sigma
}\bm{Z}_{abcde}\right]  .
\end{eqnarray}
Second, the algebra has \emph{two} ``AdS-boosts generators'' for the
\emph{same} Lorentz algebra,%
\begin{equation}%
\begin{tabular}
[c]{l}%
\begin{tabular}
[c]{ll}%
$%
\begin{array}
[c]{l}%
\left[  \bm{P}_{a},\bm{P}_{b}\right]  =\bm{J}_{ab},\\
\left[ \bm{J}^{ab}, \bm{P}_{c} \right] = \delta_{ec}^{ab} \bm{P}^{e},
\end{array}
$ & $%
\begin{array}
[c]{l}%
\left[  \bm{P}_{a}^{\prime},\bm{P}_{b}^{\prime}\right]
=\bm{J}_{ab},\\
\left[ \bm{J}^{ab}, \bm{P}_{c}^{\prime} \right] = \delta_{ec}^{ab} \bm{P}^{\prime e},
\end{array}
$%
\end{tabular}
\\
\multicolumn{1}{c}{$\left[  \bm{J}^{ab},\bm{J}_{cd}\right]
=\delta_{ecd}^{abf}\bm{J}_{\phantom{e} f}^{e}$}\\
\multicolumn{1}{c}{$\left[  \bm{P}_{a},\bm{P}_{b}^{\prime
}\right]  =\bm{Z}_{ab}.$}%
\end{tabular}
\end{equation}

The ``charges'' $\bm{Z}_{ab}$, $\bm{Z}_{a_{1} \cdots a_{5}}$ and $\bm{Z}_{a_{1} \cdots a_{5}}^{\prime}$ are Lorentz tensors, but they are not abelian,
\begin{equation}
\left[  \bm{Z}^{ab},\bm{Z}_{cd}\right]  =\delta_{ecd}%
^{abf}\bm{J}_{\phantom{e} f}^{e},
\end{equation}%
\begin{eqnarray}
\left[  \bm{Z}^{a_{1}\cdots a_{5}},\bm{Z}_{b_{1}\cdots b_{5}%
}\right]   &  = & \left[  \bm{Z}^{\prime a_{1}\cdots a_{5}}%
,\bm{Z}_{b_{1}\cdots b_{5}}^{\prime}\right] \nonumber\\
&  = & \eta^{\left[  a_{1}\cdots a_{5}\right]  \left[  c_{1}\cdots c_{5}\right]
}\varepsilon_{c_{1}\cdots c_{5}b_{1}\cdots b_{5}e}\bm{P}^{\prime
e}+\delta_{db_{1}\cdots b_{5}}^{a_{1}\cdots a_{5}e}\bm{J}_{\phantom{d} e}^{d}+\nonumber\\
&  &  -\frac{1}{3!3!5!}\varepsilon_{c_{1}\cdots c_{11}}\delta_{d_{1}d_{2}%
d_{3}b_{1}\cdots b_{5}}^{a_{1}\cdots a_{5}c_{4}c_{5}c_{6}}\eta^{\left[
c_{1}c_{2}c_{3}\right]  \left[  d_{1}d_{2}d_{3}\right]  }\bm{Z}%
^{\prime c_{7}\cdots c_{11}}.
\end{eqnarray}

This algebra also presents a behavior similar to that of the D'Auria--Fr\'{e} superalgebras; namely, the
commutators between the generators $\bm{P}_{a}$, $\bm{Z}_{ab}%
$, $\bm{Z}_{a_{1}\cdots a_{5}}$ and a Fermionic generator
$\bm{Q}$ are $\bm{Q}^{\prime}$-valued; but in contrast to
(\ref{pqp})--(\ref{z5qp}), their commutator with $\bm{Q}^{\prime}$ is
$\bm{Q}$-valued rather than zero. In this regard, the generators
$\bm{J}_{ab},\bm{Z}_{a_{1}\cdots a_{5}}^{\prime},$
$\bm{P}_{a}^{\prime}$ have a block-diagonal form on the subspace
$\left(  \bm{Q},\bm{Q}^{\prime}\right)  $; their commutator
with $\bm{Q}$ is $\bm{Q}$-valued and the one with
$\bm{Q}^{\prime}$ is $\bm{Q}^{\prime}$-valued.

We have seen several examples showing how, starting from only one original
algebra $\mathfrak{g}$ and using different semigroups, different resonant
subalgebras can arise (see Theorem~\ref{ResSubAlgTh}). This is particularly
interesting if one considers the strong similarities between the semigroups
considered, which nevertheless lead to different resonant structures.

\section{\label{SecOther}Reduced Algebras of a Resonant Subalgebra}

In previous sections we have seen how information on the subspace structure of the original algebra $\mathfrak{g}$ can be used in order to find resonant subalgebras of the $S$-expanded algebra $S \times \mathfrak{g}$. In this section we shall examine how this information can be put to use in a different way, namely, by extracting \emph{reduced algebras} (in the sense of Def.~\ref{def:foralg}) from the resonant subalgebra. It is following this path that, e.g., the generalized \.{I}n\"{o}n\"{u}--Wigner contraction fits within the present scheme.

The following general theorem provides necessary conditions under which a
reduced algebra can be extracted from a resonant subalgebra.

\begin{theorem}
\label{ReducedAlgTh}
Let $\mathfrak{G}_{\mathrm{R}} = \bigoplus_{p \in I} S_{p} \times V_{p}$ be a resonant subalgebra of $\mathfrak{G}=S\times
\mathfrak{g}$, i.e., let eqs.~(\ref{VpVq=Vr}) and~(\ref{SpSq=Sr}) be satisfied.
Let $S_{p}=\hat{S}_{p}\cup \check{S}_{p}$ be a partition of the subsets
$S_{p}\subset S$ such that
\begin{eqnarray}
\hat{S}_{p} \cap \check{S}_{p} & = & \varnothing , \label{ReducedConditionSpnSq=0} \\
\check{S}_{p} \cdot \hat{S}_{q} & \subset & \bigcap_{r \in i_{\left( p,q \right)}} \hat{S}_{r}. \label{ReducedCondition SpSq=nSr}%
\end{eqnarray}

Conditions~(\ref{ReducedConditionSpnSq=0}) and~(\ref{ReducedCondition SpSq=nSr})
induce the decomposition $\mathfrak{G}_{\mathrm{R}}=\check{\mathfrak
{G}}_{\mathrm{R}}\oplus\hat{\mathfrak{G}}_{\mathrm{R}}$ on
the resonant subalgebra, where%
\begin{eqnarray}
\check{\mathfrak{G}}_{\mathrm{R}} & = & \bigoplus_{p \in I} \check{S}_{p} \times V_{p}, \\
\hat{\mathfrak{G}}_{\mathrm{R}} & = & \bigoplus_{p \in I} \hat{S}_{p} \times V_{p}.
\end{eqnarray}

When the conditions (\ref{ReducedConditionSpnSq=0})--(\ref{ReducedCondition SpSq=nSr}) hold, then
\begin{equation}
\left[ \check{\mathfrak{G}}_{\mathrm{R}}, \hat{\mathfrak{G}}_{\mathrm{R}} \right] \subset \hat{\mathfrak{G}}_{\mathrm{R}},
\end{equation}
and therefore $\left\vert \check{\mathfrak{G}}_{\mathrm{R}} \right\vert$ corresponds to a reduced algebra of $\mathfrak{G}_{\mathrm{R}}$.
\end{theorem}

\begin{proof}
Let $\check{W}_{p}=\check{S}_{p}\times V_{p}$ and $\hat{W}_{p}=\hat{S}%
_{p}\times V_{p}$. Then, using condition~(\ref{ReducedCondition SpSq=nSr}), we have%
\begin{eqnarray*}
\left[ \check{W}_{p}, \hat{W}_{q} \right] & \subset & \left( \check{S}_{p} \cdot \hat{S}_{q} \right) \times \left[ V_{p}, V_{q} \right] \\
& \subset & \bigcap_{s \in i_{\left( p, q \right)}} \hat{S}_{s} \times \bigoplus_{r \in i_{\left( p, q \right)}} V_{r} \\
& \subset & \bigoplus_{r \in i_{\left( p, q \right)}} \left[ \bigcap_{s \in i_{\left( p, q \right)}} \hat{S}_{s} \right] \times V_{r}.
\end{eqnarray*}

For each $r\in i_{\left(  p,q\right)  }$ we have $\bigcap_{s \in i_{\left( p, q \right)}} \hat{S}_{s} \subset \hat{S}_{r}$, so that
\begin{eqnarray*}
\left[ \check{W}_{p}, \hat{W}_{q} \right] & \subset & \bigoplus_{r \in i_{\left( p, q \right)}} \hat{S}_{r} \times V_{r}\\
& \subset & \bigoplus_{r \in i_{\left( p, q \right)}} \hat{W}_{r}.
\end{eqnarray*}

Since $\check{\mathfrak{G}}_{\mathrm{R}} = \bigoplus_{p \in I} \check{W}_{p}$ and $\hat{\mathfrak{G}}_{\mathrm{R}} = \bigoplus \limits_{p \in I} \hat{W}_{p}$, we finally find
\[
\left[ \check{\mathfrak{G}}_{\mathrm{R}}, \hat{\mathfrak{G}}_{\mathrm{R}} \right] \subset \hat{\mathfrak{G}}_{\mathrm{R}}
\]
and therefore $\left\vert \check{\mathfrak{G}}_{\mathrm{R}}\right\vert
$ is a reduced algebra of $\mathfrak{G}_{\mathrm{R}}$.
\end{proof}

Using the structure constants~(\ref{ResonantStructureConstants}) for the
resonant subalgebra, it is possible to find the structure constants for the
reduced algebra $\left\vert \check{\mathfrak{G}}_{\mathrm{R}}\right\vert
,$%
\begin{equation}
C_{\left( a_{p}, \alpha_{p} \right) \left( b_{q}, \beta_{q} \right)}
^{\phantom{\left( a_{p}, \alpha_{p} \right) \left( b_{q}, \beta_{q} \right)} \left( c_{r}, \gamma_{r} \right)} =
K_{\alpha_{p} \beta_{q}}^{\phantom{\alpha_{p} \beta_{q}} \gamma_{r}}
C_{a_{p} b_{q}}^{\phantom{a_{p} b_{q}} c_{r}},
\text{ with } \alpha_{p}, \beta_{q}, \gamma_{r} \text{ such that }
\lambda_{\alpha_{p}} \in \check{S}_{p}, \lambda_{\beta_{q}} \in \check{S}_{q}, \lambda_{\gamma_{r}} \in \check{S}_{r}.
\end{equation}

It might be worth to notice that, when every $S_{p}\subset S$ of a resonant subalgebra includes the zero element $0_{S}$, the choice $\hat{S}_{p} = \left\{ 0_{S} \right\}$ automatically satisfies conditions (\ref{ReducedConditionSpnSq=0})--(\ref{ReducedCondition SpSq=nSr}). As a consequence, the $0_{S}$-reduction introduced in Def.~\ref{def:foralg} can be regarded as a particular case of Theorem~\ref{ReducedAlgTh}.

\subsection{Reduction of Resonant Subalgebras, WW conditions and the IW contraction}

Theorem~\ref{ReducedAlgTh} above will be useful in order to recover Theorem~3 from Ref.~\cite{Azcarraga Et Al} in this context.

Consider the resonant subalgebra from sec.~\ref{Sec W-W ResSubAlg} and the following $S_{p}$ partition, which satisfies~(\ref{ReducedConditionSpnSq=0}):
\begin{eqnarray}
\check{S}_{p} &  = & \left\{  \lambda_{\alpha_{p}}\text{, such that }\alpha
_{p}=p,\ldots,N_{p}\right\}  , \label{spdown}\\
\hat{S}_{p} &  = & \left\{  \lambda_{\alpha_{p}}\text{, such that }\alpha
_{p}=N_{p}+1,\ldots,N+1\right\}  . \label{spup}
\end{eqnarray}
In Appendix~\ref{appX} it is shown that the reduction condition~(\ref{ReducedCondition SpSq=nSr}) on (\ref{spdown})--(\ref{spup}) is equivalent to the following requirement on the $N_{p}$'s:
\begin{equation}
N_{p+1}=\left\{
\begin{array}[c]{l}
N_{p}\text{ or}\\
H_{N+1}\left(  N_{p}+1\right)
\end{array}
\right.  .
\end{equation}

\begin{figure}
\includegraphics[width=\columnwidth]{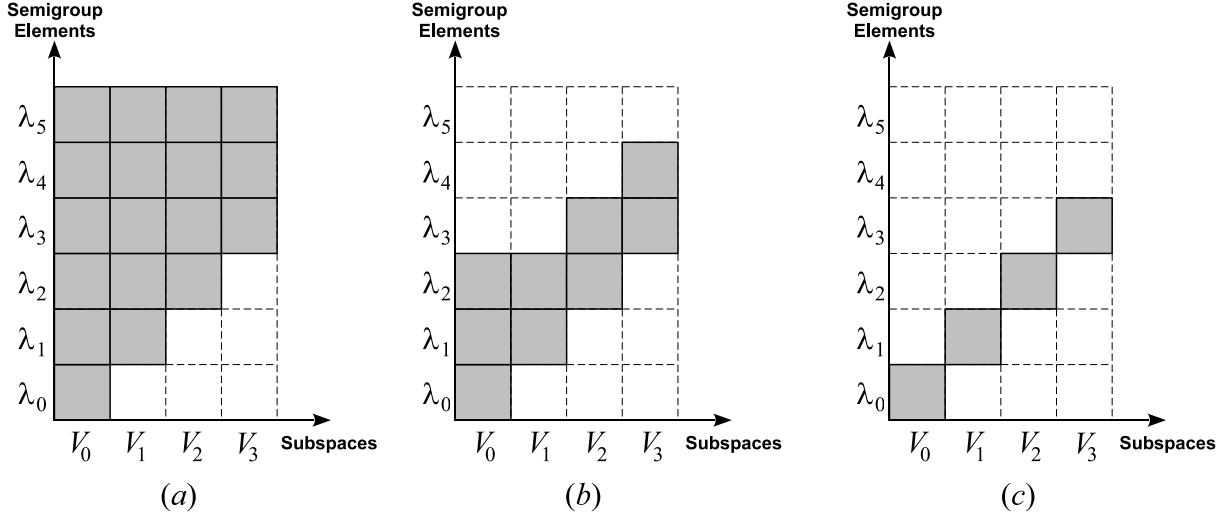}
\caption{\label{FigZamponhaForzIW}(\textit{a}) $S_{\mathrm{E}}^{(4)}$ resonant subalgebra when $\mathfrak{g} = V_{0} \oplus V_{1} \oplus V_{2} \oplus V_{3}$ satisfies the Weimar-Woods conditions. (\textit{b}) One possible reduction of the resonant subalgebra from~(\textit{a}), with $N_{0}=2$, $N_{1}=2$, $N_{2}=3$, $N_{3}=4$. (\textit{c}) Generalized \.{I}n\"{o}n\"{u}--Wigner contraction, corresponding to a different reduction of the same resonant subalgebra, with $N_{p}=p$, $p=0,1,2,3$.}
\end{figure}

This condition is exactly the one obtained in Theorem~3 of Ref.~\cite{Azcarraga Et Al}, requiring that the expansion in the Maurer--Cartan forms closes. In the $S$-expansion context the case $N_{p+1}=N_{p}=N+1$ for each $p$ corresponds to the resonant subalgebra, and the case $N_{p+1}=N_{p}=N$ to its $0_{S}$-reduction. Fig.~\ref{FigZamponhaForzIW} shows two different reductions for a resonant subalgebra where $\mathfrak{g}$ satisfies the Weimar-Woods conditions.

As stated in Ref.~\cite{Azcarraga Et Al}, the generalized \.{I}n\"{o}n\"{u}--Wigner contraction corresponds to the case $N_{p}=p$; this means that the generalized \.{I}n\"{o}n\"{u}--Wigner contraction does \emph{not} correspond to a resonant subalgebra but to its reduction. This is an important point, because, as we shall see in sec.~\ref{SecInvTen}, we have been able to define non-trace invariant tensors for resonant subalgebras and $0_{S}$-reduced algebras, but not for general reduced algebras.

As an explicit example of the application of Theorem~\ref{ReducedAlgTh}, the
$d=11$ five-brane Superalgebra is derived as a reduced algebra in the next section.

\subsection{Five-brane Superalgebra as a Reduced Algebra}

Let us recall the resonant subalgebra used in order to get the M algebra in
sec.~\ref{SecMAlgResSub}. For this case, the resonant partition
$S_{\mathrm{E}}^{\left(  2\right)  }=S_{0}\cup S_{1}\cup S_{2}$
corresponds to the one from eq.~(\ref{ResSuperPartition Sp}) for the case
$N=2$, i.e.,
\begin{eqnarray}
S_{0} &  = & \left\{  \lambda_{0},\lambda_{2},\lambda_{3}\right\}  ,\\
S_{1} &  = & \left\{  \lambda_{1},\lambda_{3}\right\}  ,\\
S_{2} &  = & \left\{  \lambda_{2},\lambda_{3}\right\}  .
\end{eqnarray}

In order to construct a reduced algebra, perform a partition of the sets
$S_{p}$ themselves, $S_{p}=\hat{S}_{p}\cup\check{S}_{p}$, such as
\begin{eqnarray}
\check{S}_{0} & = & \left\{ \lambda_{0} \right\}, \qquad
\hat{S}_{0} = \left\{ \lambda_{2}, \lambda_{3} \right\} , \\
\check{S}_{1} & = & \left\{ \lambda_{1} \right\}, \qquad
\hat{S}_{1} = \left\{ \lambda_{3} \right\} , \\
\check{S}_{2} & = & \left\{ \lambda_{2} \right\}, \qquad
\hat{S}_{2} = \left\{ \lambda_{3} \right\} .
\end{eqnarray}

\begin{figure}
\includegraphics[width=\columnwidth]{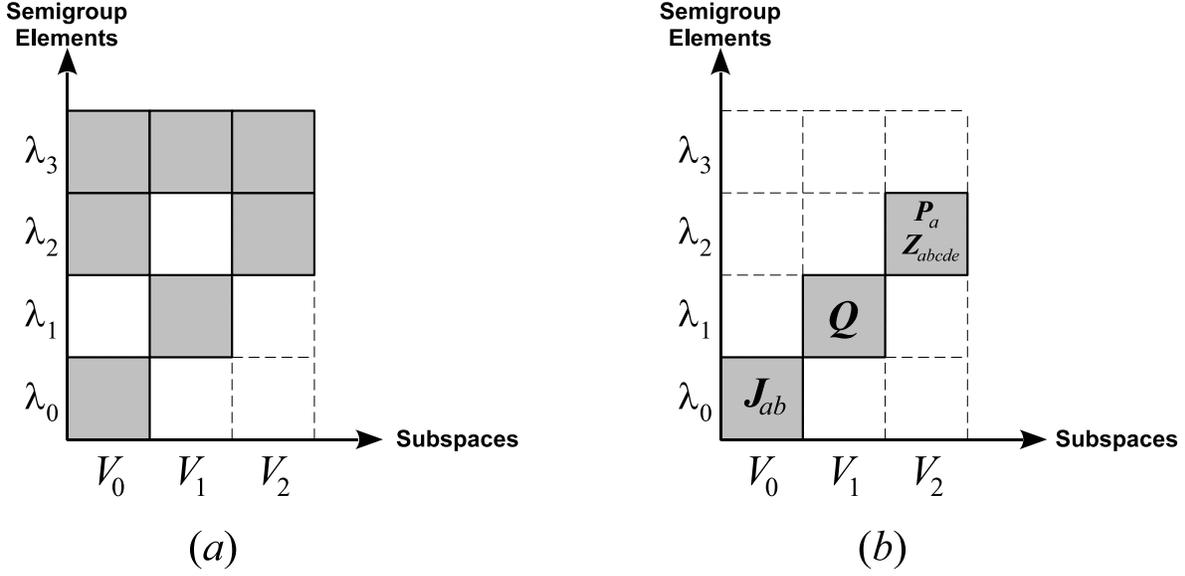}
\caption{\label{Fig Reduced5Brane}(\textit{a}) Resonant subalgebra of the $S$-expanded algebra $S_{\mathrm{E}}^{(2)} \times \mathfrak{osp} \left( 32|1 \right)$. (\textit{b}) One particular reduction of this resonant subalgebra reproduces the five-brane Superalgebra.}
\end{figure}

It is not hard to see that this partition of $S_{p}$ satisfies the reduction
conditions~(\ref{ReducedConditionSpnSq=0})--(\ref{ReducedCondition SpSq=nSr}).
For each $p$, $\hat{S}_{p}\cap\check{S}_{p}=\varnothing$, and using the
multiplication law~(\ref{Normal Expansion Product}), we have%
\begin{eqnarray*}
\check{S}_{0}\cdot\hat{S}_{0} &  \subset & \hat{S}_{0},\\
\check{S}_{0}\cdot\hat{S}_{1} &  \subset & \hat{S}_{1},\\
\check{S}_{0}\cdot\hat{S}_{2} &  \subset & \hat{S}_{2},\\
\check{S}_{1}\cdot\hat{S}_{1} &  \subset & \hat{S}_{0}\cap\hat{S}_{2},\\
\check{S}_{1}\cdot\hat{S}_{2} &  \subset & \hat{S}_{1},\\
\check{S}_{2}\cdot\hat{S}_{2} &  \subset & \hat{S}_{0}\cap\hat{S}_{2},
\end{eqnarray*}
[Compare with equations~(\ref{SuperS0S0=S0})--(\ref{SuperS2S2=S0nS2})
and~(\ref{SuperV0V0=V0})--(\ref{SuperV2V2=V0+V2})]. Therefore, we have
$\check{\mathfrak{G}}_{\mathrm{R}}=\left(  \check{S}_{0}\times
V_{0}\right)  \oplus\left(  \check{S}_{1}\times V_{1}\right)  \oplus\left(
\check{S}_{2}\times V_{2}\right)  $, which is represented in
Fig.~\ref{Fig Reduced5Brane}, and the explicit reduced algebra $\left\vert
\check{\mathfrak{G}}_{\mathrm{R}}\right\vert $%
\begin{eqnarray}
\left[  \bm{P}_{a},\bm{P}_{b}\right]   &  = & \bm{0},\\
\left[ \bm{J}^{ab}, \bm{P}_{c} \right] & = & \delta_{ec}^{ab} \bm{P}^{e},\\
\left[  \bm{J}^{ab},\bm{J}_{cd}\right]   &  = & \delta
_{ecd}^{abf}\bm{J}_{\phantom{e} f}^{e},
\end{eqnarray}%
\begin{eqnarray}
\left[  \bm{P}_{a},\bm{Z}_{b_{1}\cdots b_{5}}\right]   &
= & \bm{0},\\
\left[ \bm{J}^{ab}, \bm{Z}_{c_{1}\cdots c_{5}} \right] & = & \frac{1}{4!}\delta_{dc_{1} \cdots c_{5}}^{abe_{1} \cdots e_{4}} \bm{Z}_{\phantom{d} e_{1} \cdots e_{4}}^{d},\\
\left[  \bm{Z}^{a_{1}\cdots a_{5}},\bm{Z}_{b_{1}\cdots b_{5}%
}\right]   &  = & \bm{0},
\end{eqnarray}%
\begin{eqnarray}
\left[  \bm{P}_{a},\bm{Q}\right]   &  = & \bm{0},\\
\left[  \bm{J}_{ab},\bm{Q}\right]   &  = & -\frac{1}{2}%
\Gamma_{ab}\bm{Q},\\
\left[  \bm{Z}_{abcde},\bm{Q}\right]   &  = & \bm{0},
\end{eqnarray}%
\[
\left\{  \bm{Q}^{\rho},\bm{Q}^{\sigma}\right\}  =-\frac
{1}{2^{3}}\left[  \left(  \Gamma^{a}C^{-1}\right)  ^{\rho\sigma}%
\bm{P}_{a}+\frac{1}{5!}\left(  \Gamma^{abcde}C^{-1}\right)
^{\rho\sigma}\bm{Z}_{abcde}\right]  .
\]

This is the five-brane Superalgebra~\cite{vanHo82,deAz89}.

\section{\label{SecInvTen}Invariant Tensors for $S$-Expanded Algebras}

Finding all invariant tensors for an arbitrary algebra remains as an open
problem until now. This is not only an important mathematical problem, but
also a physical one, because an invariant tensor is a key ingredient in the
construction of Chern--Simons and Transgression forms (see, e.g.,
Refs.~\cite{Cha89,Cha90,Mora2,Polaco1,Zan05,Polaco2,Nosotros,Mora1,CECS,Nosotros2}%
), which can be used as gauge Lagrangians for a given symmetry group in an
arbitrary odd dimension. The choice of invariant tensor shapes the theory to a great extent.

A standard procedure in order to obtain an invariant tensor is to use the
(super)trace in some matrix representation of the generators of the algebra.
However, this procedure has an important limitation for $0_{S}$-reduced
algebras and, for this reason, theorems providing nontrivial invariant
tensors for $S$-expanded algebras are worth considering.

\begin{theorem}
\label{TensorInvTh}
Let $S$ be an abelian semigroup, $\mathfrak{g}$ a Lie
(super)algebra of basis $\left\{  \bm{T}_{A}\right\}  $, and let
$\left\langle \bm{T}_{A_{1}}\cdots\bm{T}_{A_{n}}\right\rangle $ be
an invariant tensor for $\mathfrak{g}$. Then, the expression%
\begin{equation}
\left\langle \bm{T}_{\left(  A_{1},\alpha_{1}\right)  }\cdots
\bm{T}_{\left(  A_{n},\alpha_{n}\right)  }\right\rangle =\alpha_{\gamma
}K_{\alpha_{1}\cdots\alpha_{n}}^{\phantom{\alpha_{1}\cdots\alpha_{n}} \gamma}\left\langle \bm{T}%
_{A_{1}}\cdots\bm{T}_{A_{n}}\right\rangle \label{InvTensorS-Exp}%
\end{equation}
where $\alpha_{\gamma}$ are arbitrary constants and $K_{\alpha_{1}\cdots\alpha_{n}}^{\phantom{\alpha_{1}\cdots\alpha_{n}} \gamma}$ is the $n$-selector for $S$, corresponds to an invariant
tensor for the $S$-expanded algebra $\mathfrak{G}=S\times\mathfrak{g}$.
\end{theorem}

\begin{proof}
The invariance condition for $\left\langle \bm{T}_{A_{1}}\cdots
\bm{T}_{A_{n}}\right\rangle $ under $\mathfrak{g}$ reads%
\begin{equation}
\sum_{p=1}^{n}X_{A_{0}\cdots A_{n}}^{\left(  p\right)  }=0, \label{InvCond g}%
\end{equation}
where%
\begin{equation}
X_{A_{0} \cdots A_{n}}^{\left( p \right)} = \left( -1 \right)
^{\mathfrak{q} \left( A_{0} \right) \left( \mathfrak{q} \left( A_{1} \right) + \cdots + \mathfrak{q} \left( A_{p-1} \right) \right)}
C_{A_{0} A_{p}}^{\phantom{A_{0} A_{p}} B} \left\langle \bm{T}_{A_{1}} \cdots \bm{T}_{A_{p-1}} \bm{T}_{B} \bm{T}_{A_{p+1}} \cdots \bm{T}_{A_{n}} \right\rangle .
\label{Xp of g}
\end{equation}

Define now
\begin{eqnarray*}
X_{\left(  A_{0},\alpha_{0}\right)  \cdots\left(  A_{n},\alpha_{n}\right)
}^{\left(  p\right)  }  &  = & \left(  -1\right)  ^{\mathfrak{q}\left(
A_{0},\alpha_{0}\right)  \left(  \mathfrak{q}\left(  A_{1},\alpha_{1}\right)
+\cdots+\mathfrak{q}\left(  A_{p-1},\alpha_{p-1}\right)  \right)  } C_{\left( A_{0}, \alpha_{0} \right) \left( A_{p}, \alpha_{p} \right)}^{\phantom{\left( A_{0}, \alpha_{0} \right) \left( A_{p}, \alpha_{p} \right)}\left( B, \beta \right)} \times\\
&  & \times \left\langle \bm{T}%
_{\left(  A_{1},\alpha_{1}\right)  }\cdots\bm{T}_{\left(
A_{p-1},\alpha_{p-1}\right)  }\bm{T}_{\left(  B,\beta\right)
}\bm{T}_{\left(  A_{p+1},\alpha_{p+1}\right)  }\cdots\bm{T}%
_{\left(  A_{n},\alpha_{n}\right)  }\right\rangle .
\end{eqnarray*}

Using the fact that $\mathfrak{q}\left(  A,\alpha\right)  =\mathfrak{q}\left(
A\right)  $ and replacing the expressions~(\ref{C=KC}) for the $S$-expansion
structure constants and~(\ref{InvTensorS-Exp}) for $\left\langle \bm{T}%
_{\left(  A_{1},\alpha_{1}\right)  }\cdots\bm{T}_{\left(  A_{n}%
,\alpha_{n}\right)  }\right\rangle $, we get
\begin{equation*}
X_{\left( A_{0}, \alpha_{0} \right) \cdots \left( A_{n}, \alpha_{n} \right)}^{\left( p \right)} = \alpha_{\gamma}
K_{\alpha_{0} \cdots \alpha_{n}}^{\phantom{\alpha_{0} \cdots \alpha_{n}} \gamma}
X_{A_{0} \cdots A_{n}}^{\left( p \right)} .
\end{equation*}
From~(\ref{InvCond g}) one readily concludes that
\begin{equation}
\sum_{p=1}^{n}X_{\left(  A_{0},\alpha_{0}\right)  \cdots\left(  A_{n}%
,\alpha_{n}\right)  }^{\left(  p\right)  }=0. \label{InvCond Sxg}%
\end{equation}

Therefore, $\left\langle \bm{T}_{\left( A_{1}, \alpha_{1} \right)}
\cdots \bm{T}_{\left( A_{n}, \alpha_{n} \right)} \right\rangle
=\alpha_{\gamma} K_{\alpha_{1} \cdots \alpha_{n}}^{\phantom{\alpha_{1} \cdots \alpha_{n}} \gamma} \left\langle
\bm{T}_{A_{1}} \cdots \bm{T}_{A_{n}} \right\rangle$ is an invariant tensor for $\mathfrak{G}=S\times\mathfrak{g}$.
\end{proof}

It is worth to notice that, in general, the expression%
\begin{equation}
\left\langle \bm{T}_{\left(  A_{1},\alpha_{1}\right)  }\cdots
\bm{T}_{\left(  A_{n},\alpha_{n}\right)  }\right\rangle =\sum_{m=0}%
^{M}\alpha_{\gamma}^{\beta_{1}\cdots\beta_{m}} K_{\beta_{1} \cdots \beta_{m} \alpha_{1} \cdots \alpha_{n}}^{\phantom{\beta_{1} \cdots \beta_{m} \alpha_{1} \cdots \alpha_{n}} \gamma}\left\langle
\bm{T}_{A_{1}}\cdots\bm{T}_{A_{n}}\right\rangle
,\label{AlfaGammaBeta1...m}%
\end{equation}
where $M$ is the number of elements of $S$ and $\alpha_{\gamma}^{\beta
_{1}\cdots\beta_{m}}$ are arbitrary constants, is also an invariant tensor for $\mathfrak{G}=S\times\mathfrak{g}$. An example of (\ref{AlfaGammaBeta1...m}) is provided by the supertrace. As a matter of fact, when the generators $\bm{T}_{A}$ are in some matrix representation, and the generators $\bm{T}_{\left(  A,\alpha\right)  }$ in the matrix representation $\bm{T}_{\left(  A,\alpha\right)  }=\left[  \lambda_{\alpha}\right]_{\mu}^{\ \nu}\bm{T}_{A},$ with $\left[  \lambda_{\alpha}\right]_{\mu}^{\ \nu}$ given in eq.~(\ref{MatrixSemigroupRep}), we have
\begin{equation}
\operatorname*{STr}\left(  \bm{T}_{\left(  A_{1},\alpha_{1}\right)
}\cdots\bm{T}_{\left(  A_{n},\alpha_{n}\right)  }\right)
=K_{\gamma \alpha_{1} \cdots \alpha_{n}}^{\phantom{\gamma \alpha_{1} \cdots \alpha_{n}} \gamma}\operatorname*{Str}%
\left(  \bm{T}_{A_{1}}\cdots\bm{T}_{A_{n}}\right)  ,
\end{equation}
where $\operatorname*{STr}$ is the (super)trace for the $\bm{T}%
_{\left(  A,\alpha\right)  }$ generators and $\operatorname*{Str}$ the one for
the $\bm{T}_{A}$ generators.

Even though the expression~(\ref{AlfaGammaBeta1...m}) could be
regarded as more general than the one from eq.~(\ref{InvTensorS-Exp}), this is not the case. Using only the associativity and closure of the semigroup
product, it is always possible to reduce eq.~(\ref{AlfaGammaBeta1...m}) to
eq.~(\ref{InvTensorS-Exp}), which in this way turns out to be more ``fundamental.''

Given an invariant tensor for an algebra, its components valued on a subalgebra are by themselves an invariant tensor for the subalgebra (if they do not vanish). For the case of resonant subalgebras, and provided all the $\alpha_{\gamma}$'s are different from zero, the invariant tensor for the resonant subalgebra never vanishes. As matter of fact, given a resonant subset partition $S = \bigcup_{p \in I} S_{p}$, and denoting the basis of $V_{p}$ as $\left\{ \bm{T}_{a_{p}} \right\}$, the $\mathfrak{G}_{\mathrm{R}}$-valued components of~(\ref{InvTensorS-Exp}) are given by
\begin{equation}
\left\langle \bm{T}_{\left( a_{p_{1}}, \alpha_{p_{1}} \right)} \cdots \bm{T}_{\left( a_{p_{n}}, \alpha_{p_{n}} \right)} \right\rangle = \alpha_{\gamma} K_{\alpha_{p_{1}} \cdots \alpha_{p_{n}}}^{\phantom{\alpha_{p_{1}} \cdots \alpha_{p_{n}}} \gamma} \left\langle \bm{T}_{a_{p_{1}}} \cdots \bm{T}_{a_{p_{n}}} \right\rangle , \text{ with } \lambda_{\alpha_{p}} \in S_{p}
\label{InvTensorS-ExpResSubAlg}%
\end{equation}

These components form an invariant tensor for the resonant subalgebra
$\mathfrak{G}_{\mathrm{R}} = \bigoplus_{p \in I} S_{p} \times V_{p}$. Since $S$ is closed under the product~(\ref{Lambda1x...xLambda n = Lambda Gamma}), for every choice of indices $\alpha_{p_{1}}, \ldots, \alpha_{p_{n}}$ there always exists a value of $\gamma$ such that $K_{\alpha_{p_{1}} \cdots \alpha_{p_{n}}}^{\phantom{\alpha_{p_{1}} \cdots \alpha_{p_{n}}} \gamma}=1$, and therefore~(\ref{InvTensorS-ExpResSubAlg}) does not
vanish (provided that $\forall \gamma, \alpha_{\gamma} \neq0$).

However, an interesting nontrivial point is that a $0_{S}$-reduced algebra is \emph{not} a subalgebra, and therefore, in general the $0_{S}$-reduced
algebra-valued components of expressions~(\ref{InvTensorS-Exp})
or~(\ref{InvTensorS-ExpResSubAlg}) do \emph{not} lead to an invariant tensor. The following theorem offers a solution by providing a general expression for an invariant tensor for a $0_{S}$-reduced algebra.

\begin{theorem}
\label{TensorInv0FTh}
Let $S$ be an abelian semigroup with nonzero elements
$\lambda_{i},$ $i=0,\ldots,N,$ and $\lambda_{N+1}=0_{S}$. Let $\mathfrak{g}$
be a Lie (super)algebra of basis $\left\{  \bm{T}_{A}\right\}  $, and
let $\left\langle \bm{T}_{A_{1}}\cdots\bm{T}_{A_{n}}\right\rangle
$ be an invariant tensor for $\mathfrak{g}$. The expression%
\begin{equation}
\left\langle \bm{T}_{\left(  A_{1},i_{1}\right)  }\cdots\bm{T}%
_{\left(  A_{n},i_{n}\right)  }\right\rangle =\alpha_{j}
K_{i_{1} \cdots i_{n}}^{\phantom{i_{1} \cdots i_{n}} j}
\left\langle \bm{T}_{A_{1}}\cdots\bm{T}_{A_{n}%
}\right\rangle \label{InvTensor 0s-Reduced}%
\end{equation}
where $\alpha_{j}$ are arbitrary constants, corresponds to an invariant tensor for the $0_{S}$-reduced algebra obtained from $\mathfrak{G}=S\times\mathfrak{g} $.
\end{theorem}

\begin{proof}
This theorem is actually a corollary of Theorem~\ref{TensorInvTh}; imposing
$\alpha_{N+1}=0$ in Theorem~\ref{TensorInvTh} and writing the $i_{0}\cdots
i_{n}$ components of eq.~(\ref{InvCond Sxg}) one gets
\begin{equation}
\sum_{p=1}^{n}X_{\left(  A_{0},i_{0}\right)  \cdots\left(  A_{n},i_{n}\right)
}^{\left(  p\right)  }=0. \label{EXio...in=0}%
\end{equation}

Using the expressions for the $S$-expansion structure constants~(\ref{C=KC}) and for the invariant tensor~(\ref{InvTensor 0s-Reduced}), one finds
\begin{eqnarray*}
X_{\left( A_{0}, i_{0} \right) \cdots \left( A_{n}, i_{n} \right)}^{\left( p \right)} & = &
\left( -1 \right)^{\mathfrak{q} \left( A_{0}, i_{0} \right)
\left( \mathfrak{q} \left( A_{1}, i_{1} \right) + \cdots + \mathfrak{q} \left( A_{p-1}, i_{p-1} \right) \right)} \times \\
& & \times \left( K_{i_{0} i_{p}}^{\phantom{i_{0} i_{p}} k}
C_{A_{0} A_{p}}^{\phantom{A_{0} A_{p}} B}
\alpha_{j} K_{i_{1} \cdots i_{p-1} k i_{p+1} \cdots i_{n}}^{\phantom{i_{1} \cdots i_{p-1} k i_{p+1} \cdots i_{n}} j}
\left\langle \bm{T}_{A_{1}} \cdots \bm{T}_{p-1} \bm{T}_{B} \bm{T}_{p+1} \cdots \bm{T}_{A_{n}} \right\rangle + \right. \\
& & \left. + K_{i_{0} i_{p}}^{\phantom{i_{0} i_{p}} N+1}
C_{A_{0} A_{p}}^{\phantom{A_{0} A_{p}} B} \alpha_{j} K_{i_{1} \cdots i_{p-1} \left( N+1 \right) i_{p+1} \cdots i_{n}}^{\phantom{i_{1} \cdots i_{p-1} \left( N+1 \right) i_{p+1} \cdots i_{n}} j} \left\langle \bm{T}_{A_{1}} \cdots \bm{T}_{p-1} \bm{T}_{B} \bm{T}_{p+1} \cdots \bm{T}_{A_{n}} \right\rangle \right) .
\end{eqnarray*}

Since%
\[
\lambda_{i_{1}}\cdots\lambda_{i_{p-1}}\lambda_{N+1}\lambda_{i_{p+1}}%
\cdots\lambda_{i_{n}}=\lambda_{N+1},
\]
we have
\[
K_{i_{1} \cdots i_{p-1} \left( N+1 \right) i_{p+1} \cdots i_{n}}
^{\phantom{i_{1} \cdots i_{p-1} \left( N+1 \right) i_{p+1} \cdots i_{n}} j} = 0,
\]
and then,%
\begin{eqnarray*}
X_{\left(  A_{0},i_{0}\right)  \cdots\left(  A_{n},i_{n}\right)  }^{\left(
p\right)  }  &  = & \left(  -1\right)  ^{\mathfrak{q}\left(  A_{0},i_{0}\right)
\left(  \mathfrak{q}\left(  A_{1},i_{1}\right)  +\cdots+\mathfrak{q}\left(
A_{p-1},i_{p-1}\right)  \right)  } K_{i_{0}i_{p}}^{\phantom{i_{0}i_{p}} k} C_{A_{0}A_{p}}^{\phantom{A_{0}A_{p}} B} \times\\
&  &  \times \alpha_{j}
K_{i_{1} \cdots i_{p-1} k i_{p+1} \cdots i_{n}}^{\phantom{i_{1} \cdots i_{p-1} k i_{p+1} \cdots i_{n}} j}\left\langle
\bm{T}_{A_{1}}\cdots\bm{T}_{p-1}\bm{T}_{B}%
\bm{T}_{p+1}\cdots\bm{T}_{A_{n}}\right\rangle .
\end{eqnarray*}

But $K_{ij}^{\phantom{ij} k}C_{AB}^{\phantom{AB} C}$ are the structure constants [see eq.~(\ref{0-Reduced Algebra})] of the $0_{S}$-reduced algebra of $S\times\mathfrak{g}$, and therefore, from eq.~(\ref{EXio...in=0}) we find that~(\ref{InvTensor 0s-Reduced}) provides an invariant tensor for it.
\end{proof}

For the $0_{S}$-reduction of a resonant subalgebra, the proof is analogous to the one given above, and
we have that
\begin{equation}
\left\langle \bm{T}_{\left(  a_{p_{1}},i_{p_{1}}\right)  }%
\cdots\bm{T}_{\left(  a_{p_{n}},i_{n}\right)  }\right\rangle =\alpha
_{j}K_{i_{p_{1}}\cdots i_{p_{n}}}^{\phantom{i_{p_{1}}\cdots i_{p_{n}}} j}\left\langle \bm{T}%
_{i_{p_{1}}}\cdots\bm{T}_{i_{p_{n}}}\right\rangle ,\text{ such that
}\lambda_{i_{p}}\in S_{p},
\end{equation}
is an invariant tensor for the $0_{S}$-reduced algebra of $\mathfrak{G}%
_{\mathrm{R}}=\sum_{p\in I}S_{p}\times V_{p}$.

The usefulness of this theorem comes from the fact that, in general, the
(super)trace in the adjoint representation for $0_{S}$-reduced algebras can
give only a very small number of components of~(\ref{InvTensor 0s-Reduced}).

As a matter of fact, using the adjoint representation given by the $0_{S}%
$-reduced structure constants in eq.~(\ref{0-Reduced Algebra}), one finds
\begin{equation}
\operatorname*{STr}\left(  \bm{T}_{\left(  A_{1},i_{1}\right)  }%
\cdots\bm{T}_{\left(  A_{n},i_{n}\right)  }\right)  =K_{j_{1}i_{1}}^{\phantom{j_{1}i_{1}} j_{2}}K_{j_{2}i_{2}}^{\phantom{j_{2}i_{2}} j_{3}} \cdots K_{j_{n-1}i_{n-1}}^{\phantom{j_{n-1}i_{n-1}} j_{n}} K_{j_{n}i_{n}}^{\phantom{j_{n}i_{n}} j_{1}}\operatorname*{Str}\left(
\bm{T}_{A_{1}}\cdots\bm{T}_{A_{n}}\right)  ,
\end{equation}
and since $\lambda_{i}\lambda_{j}=\lambda_{k\left(  i,j\right)  }$ implies
that $\lambda_{i},\lambda_{j}\neq0_{S}$, one ends up with
\begin{equation}
\operatorname*{STr}\left(  \bm{T}_{\left(  A_{1},i_{1}\right)  }%
\cdots\bm{T}_{\left(  A_{n},i_{n}\right)  }\right)  =K_{j_{1} i_{1} \cdots i_{n}}^{\phantom{j_{1} i_{1} \cdots i_{n}} j_{1}}\operatorname*{Str}\left(  \bm{T}_{A_{1}}\cdots\bm{T}_{A_{n}}\right)  .
\end{equation}

In general, this expression has less components than
eq.~(\ref{InvTensor 0s-Reduced}). In order to see this, it is useful to analyze
the case when there is also an identity element in the semigroup, $\lambda
_{0}=e$, and each $\lambda_{i}$ appears only once in each row and each column
of the semigroup's multiplication table (i.e., for each $\lambda_{i}%
,\lambda_{j}\neq e,$ we have $\lambda_{i}\lambda_{j}\neq\lambda_{i}$ and
$\lambda_{i}\lambda_{j}\neq\lambda_{j}$). In this case,
$K_{j_{1} i_{1} \cdots i_{n}}^{\phantom{j_{1} i_{1} \cdots i_{n}} j_{1}} =
K_{i_{1} \cdots i_{n}}^{\phantom{i_{1} \cdots i_{n}} 0}$,
and the only nonvanishing component of the (super)trace is
\begin{equation}
\operatorname*{STr}\left(  \bm{T}_{\left(  A_{1},i_{1}\right)  }%
\cdots\bm{T}_{\left(  A_{n},i_{n}\right)  }\right)  =K_{i_{1} \cdots i_{n}}^{\phantom{i_{1}\cdots i_{n}} 0}\operatorname*{Str}\left(  \bm{T}_{A_{1}}%
\cdots\bm{T}_{A_{n}}\right)  ,
\end{equation}
which is clearly smaller than~(\ref{InvTensor 0s-Reduced}). In the expansion
case $S=S_{\mathrm{E}}$, we have $K_{i_{1} \cdots i_{n}}^{\phantom{i_{1} \cdots i_{n}} 0} = \delta_{H_{N+1}\left(  i_{1}+\cdots+i_{n}\right)  }^{0}=\delta_{i_{1}%
+\cdots+i_{n}}^{0}$ and therefore, the only non-vanishing component of the
(super)trace is%
\begin{equation}
\operatorname*{STr}\left(  \bm{T}_{\left(  A_{1},0\right)  }%
\cdots\bm{T}_{\left(  A_{n},0\right)  }\right)  =\operatorname*{Str}%
\left(  \bm{T}_{A_{1}}\cdots\bm{T}_{A_{n}}\right)  .
\end{equation}

The advantage of the invariant tensor~(\ref{InvTensor 0s-Reduced}) as opposed
to the (super)trace is now clear; in the case $S_{\mathrm{E}}\times
\mathfrak{g}$, the (super)trace only gives a trivial repetition of the
invariant tensor of $\mathfrak{g}$, and for a resonant subalgebra, just a
piece of it.

One last remark on the invariant tensor~(\ref{InvTensor 0s-Reduced}) is that
for the particular case $S=S_{\mathrm{E}}$, since $K_{i_{1} \cdots i_{n}}^{\phantom{i_{1} \cdots i_{n}} j}=\delta_{H_{N+1}\left(  i_{1}+\cdots+i_{n}\right)  }^{j}%
=\delta_{i_{1}+\cdots+i_{n}}^{j},$ a topological density or a Chern--Simons
form constructed using the invariant tensor~(\ref{InvTensor 0s-Reduced})
coincides with the one from Ref.~\cite{Azcarraga Et Al} for the choice
$\alpha_{\gamma}=\lambda^{\gamma}$, where $\lambda^{\gamma}$ stands for a
power of the expansion parameter of the free differential algebra.

\section{\label{SecConclusions}Conclusions}

We have discussed how one can obtain a bunch of Lie algebras starting from an original one by choosing an abelian semigroup and applying the general theorems~\ref{ResSubAlgTh} and~\ref{ReducedAlgTh}, which give us ``resonant subalgebras'' and what has been dubbed ``reduced algebras.'' This procedure is a natural outgrowth of the method of Maurer--Cartan forms power-series expansion presented in Ref.~\cite{Azcarraga Et Al}, from the point of view of the Lie algebra generators and using an arbitrary abelian semigroup. The $S$-expansion presented here has the feature of being very simple and direct; given a semigroup, one needs only solve the resonance condition~(\ref{SpSq=Sr}) in order to get a resonant subalgebra, and the very similar reduction conditions~(\ref{ReducedConditionSpnSq=0})--(\ref{ReducedCondition SpSq=nSr}) in order to get a reduced one. These have been solved in several examples in order to show how both theorems work, for general algebra structures as well as for very explicit cases, e.g., $d=11$ superalgebras. As expected, the $S$-expansion scheme reproduces exactly the results of the Maurer--Cartan forms power-series expansion for a particular choice of semigroup, but it is also possible to get interesting new results using other alternatives, as shown in sec.~\ref{SecZ4Exp}.

The examples of the $S$-expansion procedure have been chosen according to their relevance for the long-term goal of understanding the geometric formulation of 11-dimensional Supergravity. To be able to write a Lagrangian invariant under these symmetries, a key ingredient is an invariant tensor. The theorems given in sec.~\ref{SecInvTen} help fill the gap, since they go a long way beyond the simple, and sometimes trivial, invariant tensors obtained from the supertrace. Chern--Simons and Transgression forms appear as a straightforward choice for the construction of a Supergravity Lagrangian in this context~\cite{Cha89,Cha90,Mora2,Polaco1,Zan05,Polaco2,Nosotros,Mora1,CECS,Nosotros2}. In this sense, theorems~\ref{ResSubAlgTh}, \ref{ReducedAlgTh}, \ref{TensorInvTh} and \ref{TensorInv0FTh} provide a very practical ``physicist's toolbox.'' These have been used to construct a Lagrangian for the M~algebra in 11 dimensions~\cite{Iza06c} following the techniques developed in Refs.~\cite{Nosotros,Nosotros2}.

There are several ways in which this work can be extended. One of them concerns the investigation of the specific properties of the algebras generated from different choices of abelian semigroups; some kind of general classification would be particularly interesting. A first step in this direction would be the construction of the above-mentioned Lagrangians, but of course there are a lot of different possibilities to proceed. A different, and perhaps fruitful path deals with the generalization of the $S$-expansion procedure itself. The conditions of discreteness and finiteness for the semigroup have been chosen primarily for simplicity, but it seems as though they could be removed in a generalized setting. The abelianity condition, on the other hand, is essential for all of our results to hold, and it is not clear whether it could be relaxed. Removing this requirement, a set with both commuting and anticommuting elements could be considered. If this possibility turns out to be feasible (which is far from trivial; think of the Jacobi identity), it would provide a way to derive superalgebras from ordinary Lie algebras and viceversa.

\begin{acknowledgments}
The authors wish to thank J.~A.~de~Azc\'{a}rraga and M.~A.~Lled\'{o} for their warm hospitality at the Universitat de Val\`{e}ncia and for many enlightening discussions. F.~I. and E.~R. are grateful to D.~L\"{u}st for his kind hospitality at the Arnold Sommerfeld Center for Theoretical Physics in Munich, where part of this work was done. F.~I. wants to thank B.~Jurco for enlightening discussions on some of the topics covered in this paper. F.~I. and E.~R. were supported by grants from the German Academic Exchange Service (DAAD) and from the Universidad de Concepci\'{o}n (Chile). P.~S.  was supported by FONDECYT Grant 1040624 and by Universidad de Concepci\'{o}n through Semilla Grants 205.011.036-1S and 205.011.037-1S.
\end{acknowledgments}

\appendix

\section{\label{appX}Reduction when $\mathfrak{g}$ satisfies the Weimar-Woods
conditions}

In sec.~\ref{Sec W-W ResSubAlg} it was shown that, when $\mathfrak{g}$ satisfies the Weimar-Woods conditions, the partition $S_{\mathrm{E}}^{\left( N \right)} = \bigcup_{p=0}^{n} S_{p}$ with $S_{p} = \left\{ \lambda_{\alpha_{p}}, \text{ such that } \alpha_{p} = p, \ldots, N+1 \right\}$ is a resonant one. In this appendix we prove that,
when each subset $S_{p}$ is split as $S_{p} = \check{S}_{p} \cup \hat{S}_{p}$, with
\begin{eqnarray}
\check{S}_{p} & = & \left\{ \lambda_{\alpha_{p}}, \text{ such that } \alpha_{p} = p, \ldots, N_{p} \right\} , \label{W-WReducedPartitionDown} \\
\hat{S}_{p} & = & \left\{ \lambda_{\alpha_{p}}, \text{ such that } \alpha_{p} = N_{p}+1, \ldots, N+1 \right\} , \label{W-WReducedPartitionUp}
\end{eqnarray}
then the reduction condition~(\ref{ReducedCondition SpSq=nSr}) from Theorem~\ref{ReducedAlgTh} is satisfied when
\begin{equation}
N_{p+1}=\left\{
\begin{array}
[c]{l}%
N_{p}\text{ or}\\
H_{N+1}\left(  N_{p}+1\right)
\end{array}
\right.  .
\end{equation}

Before proceeding, it is worth to notice that the
partition~(\ref{W-WReducedPartitionDown})--(\ref{W-WReducedPartitionUp})
automatically satisfies the following three properties:
\begin{eqnarray}
N_{p} \geq N_{q} & \Leftrightarrow & \hat{S}_{p}\subset\hat{S}_{q},
\label{Np>Nq <--> SpUp C SqUp} \\
\check{S}_{0} \cdot \hat{S}_{q} & = & \hat{S}_{q}, \label{SoDown x SqUp = SqUp} \\
\check{S}_{p} \cdot \hat{S}_{q} & \subset & \hat{S}_{x}\text{ such that }N_{x}\leq H_{N+1}\left(  p+N_{q}\right)  . \label{SpDown x Sq Up = Sx Up}
\end{eqnarray}

Since $\mathfrak{g}$ is assumed to satisfy the Weimar-Woods
conditions~\cite{W-W,W-W2}, condition~(\ref{ReducedCondition SpSq=nSr}) now reads
\begin{equation}
\check{S}_{p} \cdot \hat{S}_{q} \subset \bigcap_{r=0}^{H_{n} \left( p+q \right)} \hat{S}_{r}, \label{W-WReducedCondition SpSq=nSr}
\end{equation}
where $\check{S}_{p}$ and $\hat{S}_{q}$ are given by
(\ref{W-WReducedPartitionDown})--(\ref{W-WReducedPartitionUp}).

Let us analyze this condition for the particular case $p=0$:
\begin{equation}
\check{S}_{0} \cdot \hat{S}_{q} \subset \bigcap_{r=0}^{q} \hat{S}_{r}.
\end{equation}
Using eq.~(\ref{SoDown x SqUp = SqUp}), this turns out to be equivalent to%
\begin{equation}
\hat{S}_{q} \subset \bigcap_{r=0}^{q} \hat{S}_{r}.
\end{equation}

In this way, we have that for each $0\leq r\leq q$, $\hat{S}_{q}\subset\hat
{S}_{r}$, and using eq.~(\ref{Np>Nq <--> SpUp C SqUp}), we get the equivalent
condition%
\begin{equation}
\forall r\leq q,N_{r}\leq N_{q}. \label{r<q-->Nr<Nq}%
\end{equation}

This condition and the product~(\ref{Normal Expansion Product}) now imply that%
\[
\bigcap_{r=0}^{H_{n} \left( p+q \right)} \hat{S}_{r} = \hat{S}_{H_{n} \left( p+q \right)}.
\]

Using this fact, the condition~(\ref{W-WReducedCondition SpSq=nSr}) takes the
form%
\begin{equation}
\check{S}_{p}\cdot\hat{S}_{q}\subset\hat{S}_{H_{n}\left(  p+q\right)  }.
\label{SpDown x SqUp = SHn(p+q)Up}%
\end{equation}

Using eq.~(\ref{SpDown x Sq Up = Sx Up}), one finds that, in order to satisfy
this requirement, it is enough to impose that, for each $\hat{S}_{x}$ such that
$N_{x}\leq H_{N+1}\left(  N_{q}+p\right)  $, one has $\hat{S}_{x}\subset
\hat{S}_{H_{n}\left(  p+q\right)  }$. Alternatively
[cf.~eq.~(\ref{Np>Nq <--> SpUp C SqUp})], one can write%
\begin{equation}
\forall N_{x}\leq H_{N+1}\left(  N_{q}+p\right)  ,\quad N_{H_{n}\left(
p+q\right)  }\leq N_{x}%
\end{equation}
and therefore,%
\begin{equation}
N_{H_{n}\left(  p+q\right)  }\leq H_{N+1}\left(  N_{q}+p\right)  .
\end{equation}

For $p=1,$ we have%
\[
N_{H_{n}\left(  p+1\right)  }\leq H_{N+1}\left(  N_{q}+1\right)  ,
\]
and using eq. ~(\ref{r<q-->Nr<Nq}), one finds the inequalities%
\[
N_{q}\leq N_{H_{n}\left(  q+1\right)  }\leq H_{N+1}\left(  N_{q}+1\right)  ,
\]
whose solution is%
\[
N_{q+1}=\left\{
\begin{array}
[c]{l}%
N_{q}\text{ or}\\
H_{N+1}\left(  N_{q}+1\right)
\end{array}
\right.  .
\]

This solves condition~(\ref{W-WReducedCondition SpSq=nSr}).

Therefore, we have that
\begin{equation}
\left\vert \check{\mathfrak{G}}_{\mathrm{R}} \right\vert =\bigoplus_{p=0}^{n}\check{S}_{p}\times V_{p},
\end{equation}
with $\check{S}_{p}=\left\{  \lambda_{\alpha_{p}}\text{, such that }\alpha
_{p}=p,\ldots,N_{p}\right\}  $ and
\begin{equation}
N_{p+1}=\left\{
\begin{array}
[c]{l}%
N_{p}\text{ or}\\
H_{N+1}\left(  N_{p}+1\right)
\end{array}
\right.  ,
\end{equation}
is a reduced Lie algebra with structure constants%
\begin{equation}
C_{\left( a_{p}, \alpha_{p} \right) \left( b_{q}, \beta_{q} \right)}^{\phantom{\left( a_{p}, \alpha_{p} \right) \left( b_{q}, \beta_{q} \right)} \left( c_{r}, \gamma_{r} \right)} = K_{\alpha_{p} \beta_{q}}^{\phantom{\alpha_{p} \beta_{q}} \gamma_{r}} C_{a_{p} b_{q}}^{\phantom{a_{p} b_{q}} c_{r}}, \text{ with } \alpha_{p}, \beta_{p}, \gamma_{p} = p, \ldots , N_{p} .
\end{equation}

\end{document}